\documentclass{AIAA}
\usepackage{xcolor}
\usepackage{soul}
\usepackage{cancel}
\begin{document}

\title{Investigations and Improvement of Robustness of Reduced-Order Models of Reacting Flow}

\author{Cheng Huang\footnote{Assistant Research Scientist, Aerospace Engineering, huangche@umich.edu.} and Karthik Duraisamy\footnote{Associate Professor, Aerospace Engineering, kdur@umich.edu.}}
\affiliation{University of Michigan, Ann Arbor, MI, 48109}
\author{Charles L. Merkle\footnote{Professor Emeritus, School of Aeronautics and Astronautics, merkle@purdue.edu.}}
\affiliation{Purdue University, West Lafayette, IN, 47907}

\begin{abstract}
The impact of chemical reactions on the robustness and accuracy of projection-based Reduced-Order Models (ROMs) of fluid flows is investigated. Both Galerkin and Least-Squares Petrov Galerkin ROMs are shown to be less robust in reacting flows as compared to non-reacting flows. In particular, reacting flow ROMs show a strong sensitivity to \st{the} resolution and are often unstable. To identify the main underlying causes, a representative problem that contains the essential physics encountered in typical combustion dynamics problems is chosen. Comparisons with non-reacting solutions are used to assess the impact of reactions. Investigations are focused on three potential areas of significance: 1) preservation of conservation laws; 2) loss of dissipation; and 3) existence of unphysical local phenomena. Results indicate that conservation is relatively well-controlled and the global dissipation in the ROMs is actually larger than that in the underlying CFD solutions. Spurious local phenomena are, however, highly deleterious. Specifically, the steep temperature gradients that characterize combustion can lead to oscillations in local temperatures even in the absence of reactions. Representative calculations with physics-based temperature constraints verify that eliminating such excursions results in considerable improvement in both stability and future-state prediction capability. 
\end{abstract}

\maketitle

\section*{Nomenclature}
{\renewcommand\arraystretch{1.0}
\noindent\begin{tabular}{@{}l @{\quad=\quad} l@{}}
{FOM}&{Full Order Model}\\
{ROM}&{Reduced Order Model}\\
{POD}&{Proper Orthogonal Decomposition}\\
{LSPG}&{Least Squares Petrov Galerkin}\\
{$\mathbf{Q}_p$}&{spatially discrete solution variables}\\
{$\mathbf{\hat{R}}$}&{spatially discrete residuals of full order model equations}\\
{$\Omega_{ref}(t)$}&{boundary conditions}\\
{$\bar{(\cdot)}$}&{time-averaged solutions}\\
{$(\cdot)'$}&{mean subtracted unsteady solutions}\\
{$\sigma_n$}&{the $n^{th}$ singular value}\\
{$a_n(t)$}&{the $n^{th}$ POD temporal mode}\\
{$\mathbf{\Phi}_n$}&{the $n^{th}$ POD eigen-basis}\\
{$\left<\cdot,\cdot\right>_P$}&{inner product operator}\\
{$\mathbf{f}$}&{spatially discrete residuals of reduced order model equations}\\
{$\mathbf{W}_n$}&{the $n^{th}$ test function for Least Squares Petrov-Galerkin method}\\
{$A$}&{pre-exponential factor}\\
{$k$}&{kinetic energy}\\
{$s$}&{entropy}\\
{AD}&{artificial dissipation}\\
\end{tabular}}

\section{Introduction}
\label{intro}
Computational Fluid Dynamic (CFD) modeling of reacting flow serves an important role in the investigation and understanding of combustion dynamics in aerospace propulsion systems operating at high pressure and temperature, the details of which cannot be quantitatively accessed through experiment. Accurate modeling of combustion dynamics in practical combustor design is key to improving engine performance, reducing failures and avoiding the devastating phenomena of combustion instability. Though modern computational capability has moved beyond the empirically-based design analyses of the past, high-fidelity simulations of full-scale practical combustors remain out of reach for day-to-day engineering design applications. While complete LES simulations of combustion dynamics in laboratory-scale combustors are routinely pursued,~\cite{HarvazinskiPoF}
, and representative computations in small-scale full combustors have been reported~\cite{StaffelbachLES,WolfLES,Gicquel2012}, they remain prohibitively expensive. Therefore, for simulation-based design to be a reality, accurate and efficient modeling methodologies are imperative. 

In the present work, we examine the use of model reduction techniques to achieve efficient and accurate modeling of combustion dynamics in complex geometries. Specifically, we pursue projection-based Reduced Order Models (ROMs)~\cite{Lumley1997,Graham1997,Lucia2003}
that couple Proper Orthogonal Decomposition~\cite{taira2019modal} (POD) with Galerkin and Petrov-Galerkin projection methods. Reduced-order models have proven to be efficient in reducing complex partial differential equations (PDE) to low dimensional ordinary differential equations (ODE) in non-reacting flow problems such as flow control~\cite{Barbagallo2012,Barbagallo2009,Barbagallo2011} and aeroelasticity~\cite{Lieu2007,Lucia2004}. Recent studies have extended ROMs to reacting flow problems~\cite{MunipalliRBM,Huang2007} and preliminary explorations of the POD/Galerkin technique have been carried out~\cite{Huang2014,Huang2015} using a simple 1D solver to establish a basic approach for ROM construction and its characteristics in reacting flows. More recently, the current authors initiated the first attempt to apply ROM techniques on relevant combustor flow simulations~\cite{Huang2018Jan,Huang2018July}, which identified steep temperature gradients as one of the underlying challenges of reacting flow ROM development.

This work is part of a larger initiative in which the authors are developing a Multi-Fidelity Framework using Reduced Order Modeling (ROM) to enable efficient modeling of combustion dynamics in practical full-scale engines. Using a rocket engine as an example, it should be noted that complex flow phenomena are limited to the region in and around the injector element, where the propellants mix and react through actions of shear, swirl, and/or direct impingement. In large, high-performance rocket engine combustors, where the problem of instability is particularly severe, hundreds of injector elements are typically used. The design and operation of these individual elements control the spatio-temporal aspects of mixing, reaction, and the overall heat addition field. Although the number of injector elements is large (which precludes application of high-fidelity methods to the entire combustor), the elements themselves are rather small (the largest are on the order of 1 cm). This fact can be used to an advantage by restricting the detailed characterization of the unsteady flowfield and its response to flow perturbations to one or a small set of injector elements. Then, a generalized reduced order model (ROM) for  injector elements can be extracted from the simulations and implemented in multi-fidelity design framework suitable for engineering analysis. The idea of such a multi-Fidelity framework has been successfully demonstrated in simpler model problems~\cite{Huang2019,Xu2019}.

It is well-recognized that one of the most critical issues in ROM development is the lack of robustness and stability when reducing a complex multi-scale system to a low-order ODE. As reported, the issues can come from the inherent lack of numerical stability in the POD/Galerkin method itself~\cite{Rempfer2000}, truncation of low-energy dissipative POD modes~\cite{Bergmann2009} and simplifications of higher-order equations~\cite{Noack2005}. The balanced POD technique has been proposed to build numerically stable ROM for linear systems~\cite{Rowley2005,Willcox2002}. Bergmann et al.~\cite{Bergmann2009} proposed to add residuals of the Navier-Stokes equations to account for the absence of low-energy dissipative POD modes. Moreover, Lucia et al.~\cite{Lucia2003} demonstrated the effectiveness of constructing stable ROMs by including additional artificial dissipation terms. Researchers have also tried to resolve the issues arising from the numerical properties of the system equations. Rowley et al.~\cite{Rowley2004} pointed out that defining a proper inner product can be important when dealing with model reduction of the Navier-Stokes equations. Barone et al.~\cite{Barone2008,Barone2009JCP,Barone2009report} proposed to stabilize the reduced system by symmetrizing the higher-order PDE with a preconditioning matrix. For aeroelastic applications, Amsallen and Farhat~\cite{Amsallem2014} have shown the advantages of using the descriptor form over the non-descriptor form of the governing equations. Carlberg et al.~\cite{Carlberg2017} have also demonstrated that the Least Squares Petrov-Galerkin (LSPG) method generates ROMs that are consistently more stable than those generated by the Galerkin method via symmetrization of the discrete Jacobian.

All of the above research investigations were focused on the robustness and stability of ROMs for non-reacting flow applications. The additional physical complexity arising from chemical reactions, however, further exacerbates the robustness issues that are observed in reduced-order models. Therefore, based on the challenges of previous reacting flow ROM development~\cite{Huang2018Jan,Huang2018July}, the main objectives of the present paper are to:  
\begin{enumerate}
    \item Compare ROMs for reacting and non-reacting flow conditions based upon a test problem that is representative of combustion dynamics problems;
    \item Reveal and narrow down the main underlying causes of ROM stability issues in reacting flow simulations; and
    \item Implement and evaluate potential resolutions to improve ROM characteristics.
\end{enumerate}
Although our focus is on reacting flow problems, it is anticipated that observations gained from the present work will be useful for improving ROMs for non-reacting flows also.

The remainder of the paper is organized as follows. In Section~\ref{formulation}, we briefly review the CFD model that is used to generate the datasets for ROM development and introduce the procedure for POD mode generation and the Galerkin and Least Squares Petrov Galerkin (LSPG) model reduction techniques. In Section~\ref{testprob}, we present the computational setup of the test problem used for the ROM evaluations. In Section~\ref{results}, we present computational results and assess the accuracy and robustness of ROMs developed from the benchmark test problem followed by a detailed diagnosis of the impact of combustion on ROM stability along with the implementation of a simple modification that provides immediate relief from the most severe symptoms. In the last section, we provide concluding remarks and address future work.

\section{Formulation}
\label{formulation}
\subsection{Governing Equations}
The computational infrastructure used for the full and reduced-order models solves conservation equations for mass, momentum, energy and species mass fractions in a coupled fashion and is based on an in-house CFD code that has previously been used to model combustion instabilities~\cite{HarvazinskiPoF}. A detailed description of the Full Order Model (FOM) equations can be found in Appendix~\ref{appendix:fom_eq}. The semi-discretized version of the governing equations in terms of primitive variables yields the ODE system, 
\begin{equation}
    \frac{\partial{\mathbf{Q}_p}}{\partial{t}}=\hat{\mathbf{R}}\left(\mathbf{Q}_p,\Omega_{ref}(t)\right),
    \label{cfd:discretized}
\end{equation}
where $\mathbf{Q}_p(t)=\left[Q_{p,1}(t){\ldots}Q_{p,i}(t){\ldots}Q_{p,NI}(t)\right]^T$ with $NI$ as the total number of grid points and $\mathbf{Q}_{p,i}(t)=\left[p_{i}(t),\textbf{u}_{i}(t),T_{i}(t),\textbf{Y}_{k}(t)\right]^T$ $\forall{i}$. Equation~\ref{cfd:discretized} is a high dimensional ODE composed of ($NI{\times}N_{var}$) equations where $N_{var}$ is the number of solution variables in $Q_{p,i}$. Boundary conditions are applied at boundary faces in the finite volume scheme by the explicitly specified input vector, $\Omega_{ref}(t)$. Moreover, it should be pointed out that both the FOM and ROM employ consistent finite volume discretization schemes and specifically for the current studies, second-order Roe scheme~\cite{Roe1981} with flux limiter due to Barth~\cite{barth1989} is used.

\subsection{Construction of POD bases for vector equations}
POD bases are derived from a database created by storing snapshots of Full-Order Model (FOM) solutions of Eq.~\ref{cfd:discretized} over the entire computational domain at specific time instants. To enable the ROM to faithfully predict the combustion response to various disturbances, multiple databases are created from FOM solutions, each of which is forced by driving a single (upstream or downstream) boundary condition, $\Omega_{ref}(t)$ in a periodic manner while the remaining boundary conditions are held fixed. Nonlinear effects are included by forcing at finite amplitudes. 

Given this database from FOM solutions of Eq~\ref{cfd:discretized}, POD bases are calculated based upon the vector-valued method using the unsteady part of the variables, $\mathbf{Q}'_p(t)=\mathbf{Q}_p(t)-\overline{\mathbf{Q}}_p$ (time-averaged values extracted before POD eigen-bases calculation), 
\begin{equation}
    \mathbf{Q}'_p(t)\approx\sum^{N_p}_{n=1}\hat{a}_n(t)\sigma_n\mathbf{\Phi}_n=\sum^{N_p}_{n=1}{a}_n(t)\mathbf{\Phi}_n.
    \label{pod:expansion}
\end{equation}
Here, $\sigma_n$ is the singular value and $a_n(t)$ the temporal amplitude of the $n^{th}$ POD mode of the orthonormal vector function, $\Phi_n$, where,
\begin{equation}
    \left<\mathbf{\Phi}_k,\mathbf{\Phi}_n\right>_{P}=\left\{\begin{array}{cc}
        1, & \text{if }k=n \\
        0, & \text{otherwise}
    \end{array} \right .
    \label{pod:orthogonality}
\end{equation}
with an inner product defined as $\left<\mathbf{u},\mathbf{v}\right>_{P}=\mathbf{u}^T{P}\mathbf{v}$. A normalization matrix, $P$, must be applied to scale the vector-valued POD bases. Here, we normalize all fluctuation quantities by their maximum amplitude,
\begin{equation}
    P = diag\left(\hat{P}_1\ldots\hat{P}_i\ldots\hat{P}_{NI}\right),
    \label{pod:normalization}
\end{equation}
where $\hat{P}_i=diag\left(\frac{1}{p'_{max}},\frac{1}{u'_{j,max}},\frac{1}{T'_{max}},\frac{1}{Y'_{l,max}}\right)$ and $\varphi'_{max}\equiv{\text{Max}\left\{\varphi'(x,t)\right\}}$, $\forall{x_{min}}\leq{x}\leq{x_{max}}$ and $t>0$. 

\subsection{Model Reduction}
In this work, projection-based model reduction is accomplished using both Galerkin- and Least-Squares Petrov Galerkin-POD methods. Galerkin-based model reduction procedures are formulated upon the continuous-time representation of Eq.~\ref{cfd:discretized}, which is then projected onto the $k^{th}$ test function, $\mathbf{V}_k$, chosen to be the $k^{th}$ basis, $\mathbf{\Phi}_k$, obtained in Eq.~\ref{pod:expansion}, 
\begin{equation}
    \left<\mathbf{V}_k,\frac{\partial{\mathbf{Q}_p}}{\partial{t}}\right>_P=\left<\mathbf{V}_k,\hat{\mathbf{R}}\left(\mathbf{Q}_p,\Omega_{ref}(t)\right)\right>_P.
    \label{rom:projection}
\end{equation}
where the inner product operator is introduced in Eq.~\ref{pod:orthogonality}, and $\mathbf{Q}_p$ is approximated using Eq.~\ref{pod:expansion}, following which a time-variant ODE system is obtained,
\begin{equation}
    \frac{d{\mathbf{a}(t)}}{dt}=\mathbf{f}\left(\mathbf{a}(t),\Omega_{ref}(t)\right),
    \label{rom:ode}
\end{equation}
where $\mathbf{a}(t)=\left[a_1(t)\cdots{a_i(t)}\cdots{a_{N_p}(t)}\right]^T$, $\mathbf{f}(t)=\left[f_1(t)\cdots{f_i(t)}\cdots{f_{N_p}(t)}\right]^T$, the dimension of the ROM ODE in Eq.~\ref{rom:ode} is $N_p$, orders of magnitudes smaller than the $NI\times{N_{var}}$ degrees of freedom in Eq.~\ref{cfd:discretized}. As indicated, the same boundary condition, $\Omega_{ref}(t)$, appears in the ROM ODE as in the CFD equations, Eq.~\ref{cfd:discretized}.

The Least Squares Petrov-Galerkin (LSPG) model-reduction technique differs from the Galerkin procedure in that it is formulated from an implicit, discrete-time representation of Eq.~\ref{cfd:discretized}, as opposed to the continuous-time representation for the Galerkin method~\cite{Carlberg2017}. For exemplary purposes, we consider the first-order backward Euler scheme for the temporal discretization. Extension to other implicit methods is straightforward.

The LSPG method is based upon the fully discretized algebraic version of Eq.~\ref{cfd:discretized},
\begin{equation}
    \frac{\mathbf{Q}^n_p-\mathbf{Q}^{n-1}_p}{\Delta{t}}=\hat{\mathbf{R}}\left(\mathbf{Q}^n_p,\Omega^n_{ref}\right),
    \label{cfd:time_discretized}
\end{equation}
where $n$ represents the physical time step. Following the same procedure as Eq.~\ref{rom:projection}, Eq.~\ref{cfd:time_discretized} is projected onto the $k^{th}$ test function, $\mathbf{W}^n_k$, 
\begin{equation}
    \left<\mathbf{W}^n_k,\frac{\mathbf{Q}^n_p-\mathbf{Q}^{n-1}_p}{\Delta{t}}-\hat{\mathbf{R}}\left(\mathbf{Q}^n_p,\Omega_{ref}(t)\right)\right>_P=\textbf{r}^n.
    \label{lspg-rom:projection}
\end{equation}
where the right-hand-side term, $\mathbf{r}^n$, represents the residual of the reduced system equations. 
Instead of using the same POD basis for the test function as in the Galerkin method, Eq.~\ref{rom:projection}, the goal of the LSPG method is to find a test function, $\mathbf{W}^n_k$, such that the right-hand-side residual in Eq.~\ref{lspg-rom:projection}, $\mathbf{r}^n$, is minimized. Similar to Least squares finite element methods~\cite{hughes_GLS}, this minimization provides numerical stabilization.

By solving the minimization problem, a new time-varying test function, $\mathbf{W}^n_k$, can be defined as,
\begin{equation}
    \mathbf{W}^n_k=P^{-1}\frac{\mathbf{\Phi}_k}{\Delta{t}}-\left(\frac{\partial\hat{\mathbf{R}}}{\partial\mathbf{Q}_p}\right)P^{-1}\mathbf{\Phi}_k,
    \label{lspg-rom:test_function}
\end{equation}
where $\left(\frac{\partial\hat{\mathbf{R}}}{\partial\mathbf{Q}_p}\right)$ is evaluated using the POD expansion, Eq.~\ref{pod:reconstr}~\cite{Carlberg2017}. 
Substituting Eq.~\ref{lspg-rom:test_function} into Eq.~\ref{lspg-rom:projection}, the discretized ODE system can be obtained for the LSPG method, 
\begin{equation}
    \hat{M}^n\frac{\mathbf{a}^n-\mathbf{a}^{n-1}}{\Delta{t}}=\hat{\mathbf{f}}\left(\mathbf{a}^n,\Omega^n_{ref}\right),
    \label{rom:ode-lspg}
\end{equation}
where the elements in matrix in $\hat{M}^n$ are, $\hat{m}_{k,j}=\left<\mathbf{W}^n_k,\mathbf{\Phi}_j\right>_P$ and those in $\hat{\mathbf{f}}$ are, 
$\hat{f}_k=\left<\mathbf{W}^n_k,\hat{\mathbf{R}}\left(\mathbf{Q}^n_p,\Omega^n_{ref}\right)\right>_P$ with $\mathbf{Q}_p$ approximated using Eq.~\ref{pod:reconstr}.
The symmetric nature of the   matrix, $\hat{M}^n$,  improves robustness from a global stability standpoint.

The LSPG method requires  implicit time integration, as indicated in Eqs.~\ref{cfd:time_discretized} and~\ref{rom:ode-lspg}, while the standard Galerkin method can be solved using either explicit or implicit schemes.  Applying the LSPG method to Eq.~\ref{cfd:discretized} with explicit time discretization results in a ROM ODE system identical to the Galerkin result. In the section below, we assess both Galerkin and LSPG methods in terms of ROM stability.

\section{Test Problem for Reduced Order Model Assessment}
\label{testprob}

\begin{figure}
	\centering
	\includegraphics[width=1.0\textwidth]{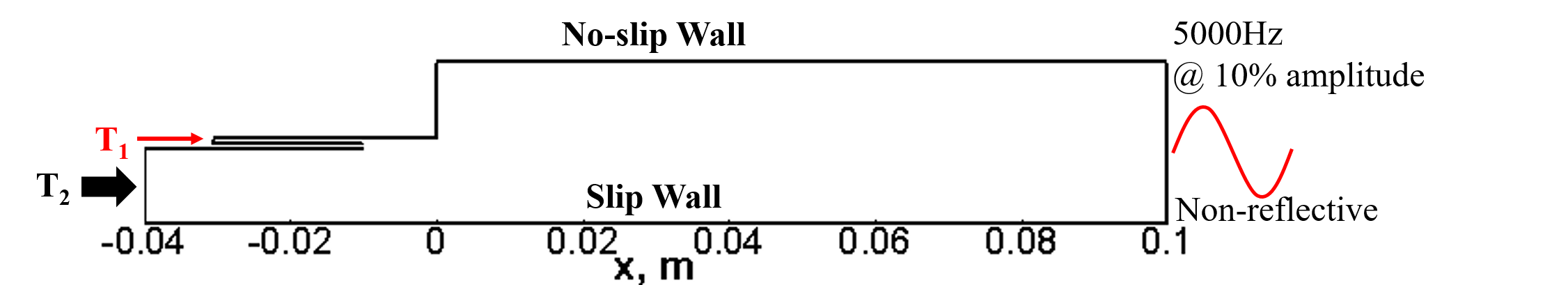}
	\caption{Overview of 2D planar rocket injector benchmark problem.}\label{geometry} 
\end{figure}

To assess the capabilities of ROMs for representing realistic combustion flowfields, a 2D-planar representation of a generic laboratory-scale combustor~\cite{YuJPP} is used. This simplified model allows ROM capabilities to be evaluated while maintaining the essential physics of interest. The specific configuration is shown in Fig.~\ref{geometry}. The problem consists of a shear coaxial injector with an outer passage, $T_1$, that introduces fuel near the downstream end of the inner passage and a coaxial center passage, $T_2$, that feeds oxidizer to the combustion chamber. The $T_1$ stream contains gaseous methane while the T2 stream is 42\% gaseous $O_2$ and 58\% gaseous $H_2O$. 

Operating conditions are maintained similar to conditions in the laboratory combustor~\cite{YuJPP,YuPhD} with an adiabatic flame temperature of approximately 2700K and an imposed chamber pressure of 1.1MPa. Both the $T_2$ and $T_1$ streams are fed with constant mass flow rates; 5.0kg/s and 0.37kg/s respectively. A non-reflective boundary condition is imposed at the downstream end to control acoustic effects on the combustion dynamics. For all the cases in the current paper, a 10\% sinusoidal perturbation at 5000Hz is imposed at the downstream boundary to generate FOM solutions for POD mode generation. Combustion is represented by the single-step global model of Westbrook and Dryer~\cite{WestbrookDryer},
\begin{equation}
    CH_4+2O_2\rightarrow{CO_2{+}2H_2{O}}.
    \label{chemistry}
\end{equation}
As reported in Ref~\cite{Huang2018Jan}, stable, accurate reconstruction of CFD solutions of flows with stiff chemistry is highly challenging. To diminish these difficulties, the present simulations are based upon a reduced pre-exponential factor, $A = 2\times{10}^{10}$, which is a factor of ten smaller than the value in~\cite{WestbrookDryer} and corresponds to a characteristic chemical time scale of approximately 0.8$\mu{s}$ and a laminar flame thickness of approximiately 1$mm$. The FOM was calculated using a constant time step at  0.1$\mu{s}$ and grid size, 0.18$mm$, in the reacting regions. Even with this reduced reaction rate, the resulting ROMs remain highly temperamental and provide a clear example of the additional difficulties engendered when reactions are present. This reduced reaction rate provides conditions more favorable for ROM development than the original stiff value while maintaining representative flame dynamics in the combustor.

\begin{figure}
	\centering
	\includegraphics[width=0.8\textwidth]{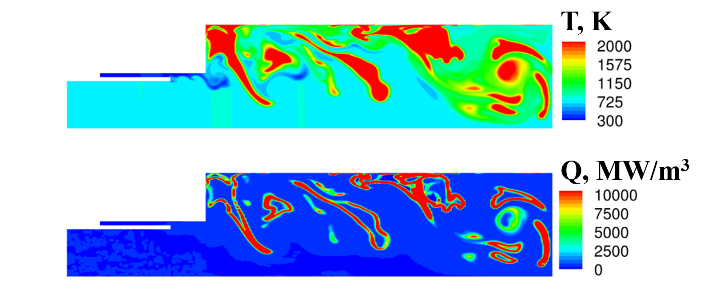}
	\caption{Representative instantaneous snapshots of temperature and heat release from CFD simulation.}\label{fom-sol} 
\end{figure}

A representative instantaneous snapshot of the reacting FOM solutions is shown in Fig.~\ref{fom-sol} to demonstrate the overall character of the flowfield and to highlight the dominant physics in the problem of interest. The combustion dynamics are characterized by highly dispersed pockets of intense heat release that are intermittently distributed in both space and time. The temperature and heat-release contours span a wide range of scales from the small eddies in the shear layers to the large-scale recirculation zone behind the dump plane. All these unique features introduce varying levels of difficulty in constructing a robust ROM.

\section{Numerical Investigations}
\label{results}

As indicated above, ROMs for reacting flows tend to be less robust than those for non-reacting flows. The goal of the paper is to identify the main underlying causes for the challenges arising in developing ROMs for reacting flows. Having identified these issues, one can then use this understanding to improve ROMs for combustion dynamics applications and potentially for non-reacting applications as well. To this end, we conduct detailed investigations on the test problem illustrated in Fig.~\ref{geometry}. To elucidate the source of ROM stability issues for reacting flows, we consider both reacting and non-reacting versions of the test problem. 

The FOM time step for all calculations was 0.1$\mu{s}$, and data were stored over a time duration of 1.0ms, corresponding to a total of 10,000 time steps. The ROM database for POD mode generation was established by down-sampling the FOM solution every 50 time steps resulting in a total of 200 POD snapshots. Clearly, both the length of the time sample and the sampling interval are important issues in any reduced-order representation. The duration of the time sample dictates the lowest frequency that can be represented by the ROM, while the sampling interval restricts the high-frequency content. The above values were chosen as a compromise to represent the dominant physics in the present test problem. A sensitivity study on the sampling interval is presented in Appendix~\ref{appendix:snapshot_effects}.

\subsection{Energy Content as a Function of the Number of POD Modes}
\label{results-energy}

\begin{figure}
	\centering
	\includegraphics[width=0.6\textwidth]{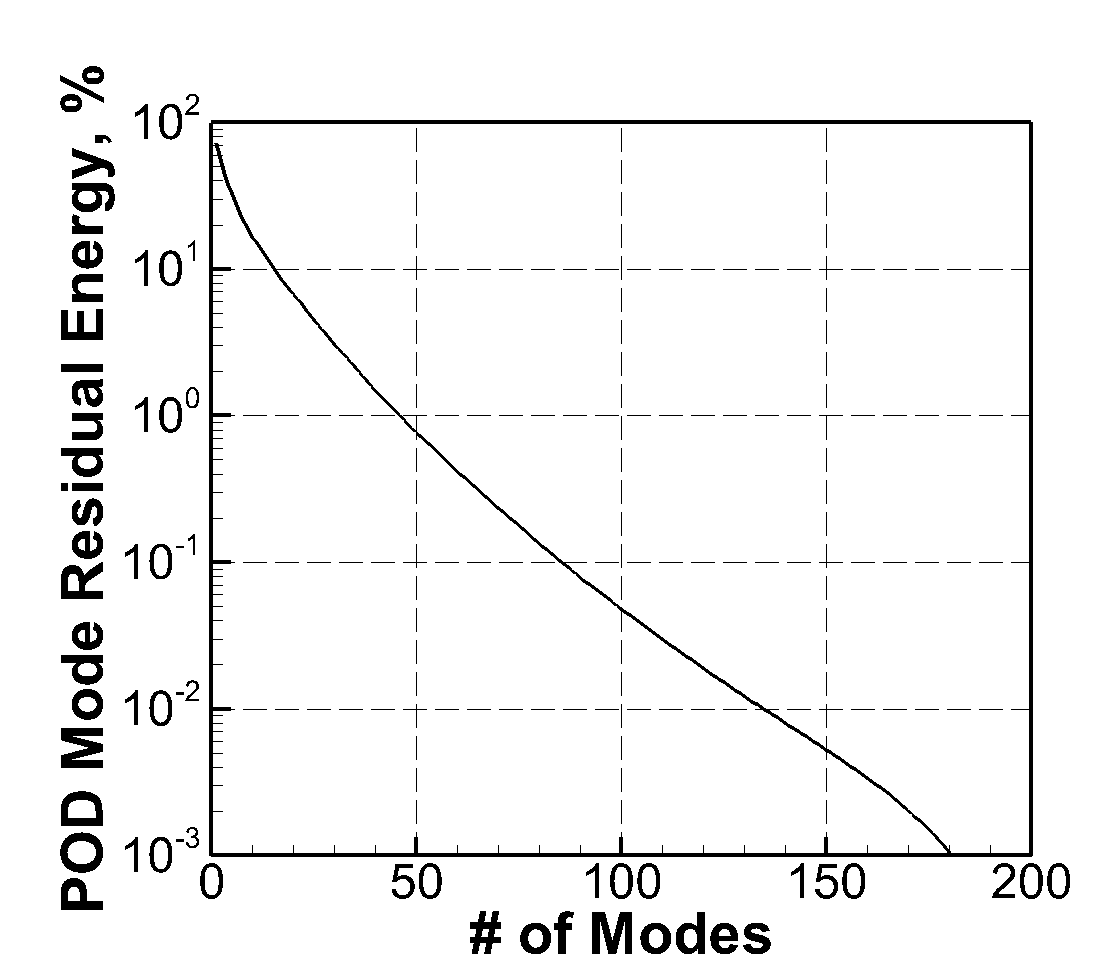}
	\caption{POD residual energy distribution from 2D rocket injector simulation.}\label{pod_res_energy} 
\end{figure}

The POD characteristics of the reacting flow problem are first investigated to understand how well the POD modes represent the original CFD dataset. The representation is based upon the POD residual energy: 
\begin{equation}
    \text{POD Residual Energy}(N_p),{\%}{=}\left(1-\frac{\sum^{N_p}_{n=1}\sigma^2_n}{\sum^{N_{p,total}}_{n=1}\sigma^2_n}\right)\times{100\%},
    \label{pod:res_energy}
\end{equation}
where $N_p$ is the number of POD modes included, and $N_{p,total}$(= 200) is the total number of snapshots in the dataset . The residual energy as a function of $N_p$ shown Fig.~\ref{pod_res_energy} reveals the amount of information omitted by the POD representation for any particular number of modes. The results show that approximately the first 15 modes must be included to capture 90\% of the total energy while 45 modes recover approximately 99\% and at least 133 modes (66.5\% of the complete set) are needed to retrieve 99.99\% of the total energy. ROM applications in the literature are often presented for problems exhibiting less physical complexity such that they are able to capture 99.9\% of the energy with a dozen or so modes. The relatively large number of modes needed here is a testimony to the wide range of dominant scales in the present problem as noted in Fig.~\ref{fom-sol}.

\subsection{ROM Reconstruction Errors as a Function of the Number of POD Modes}
\label{results-RomReconErr}

The above results illustrate the number of modes required to represent a given fraction of the energy in the original CFD solution. In the present section, we examine the accuracy with which ROM solutions can reproduce the original signal. Results are presented for ROMs based on both Galerkin and least squares Petrov-Galerkin (LSPG) projection. Galerkin schemes are commonly employed in reduced-order modeling, but it is well-known that they are susceptible to instability in non-symmetric systems, especially when details are under-resolved. The symmetrized ODE system provided by the LSPG method should provide stability improvements and more robust ROMs. Such improvements have been demonstrated for non-reacting flows~\cite{Carlberg2017}. Here, we provide a three-way comparison of the methods: the Galerkin procedure evaluated for both explicit (RK-4) and implicit time marching of the ODE system, and the LSPG system with implicit integration. (As noted above, with explicit integration the LSPG and Galerkin systems become identical.) 

\begin{figure}
	\centering
	\includegraphics[width=1.0\textwidth]{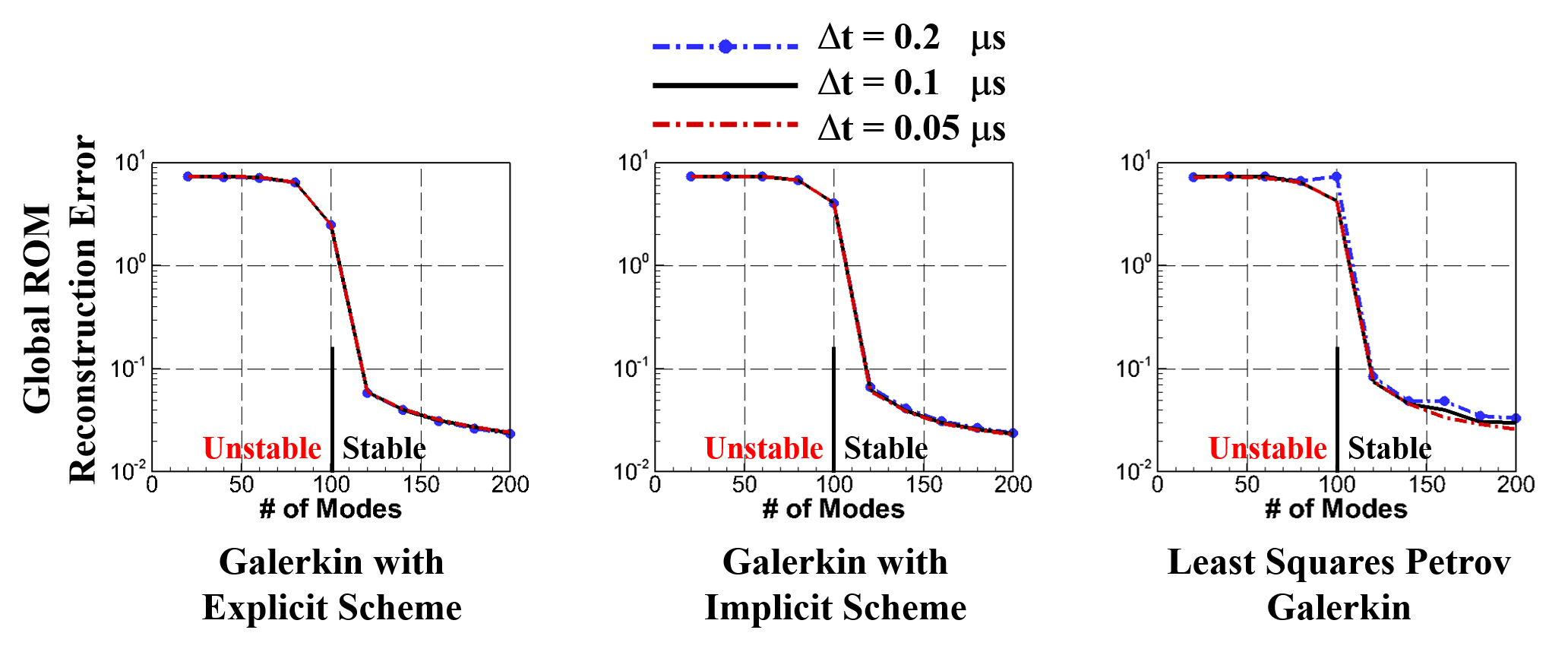}
	\caption{Global ROM reconstruction error comparisons between different projection-based ROM methods with varied computing time steps for the 2D benchmark reacting flow simulation.}\label{rom_err_vs_dt} 
\end{figure}

Comparisons of the three methods are summarized in Fig.~\ref{rom_err_vs_dt} in terms of the ROM reconstruction error,
\begin{equation}
    \text{Global ROM Reconstruction Error}(n_p)=\frac{1}{n_{nvar}}\sum^{n_{var}}_{j=1}\left(\frac{1}{NI}\sum^{NI}_{i=1}\left\|1-\frac{q'_{rom,j,i}(t)}{q'_{pod,j,i}(t)}\right\|_2\right)
    \label{rom:reconstr_err}
\end{equation}
where  $q'_{pod,j,i}(t)=P^{-1}\sum^{n_p}_{k=1}a_{pod}(t)\phi_{k,j,i}$ represent the particular solution variable. Note the errors in all $n_{var}$ variables are computed independently, summed and averaged. To ensure time step does not have a major effect, all three ROM versions are temporally integrated with three different time steps, 0.2, 0.1 and 0.05 $\mu{s}$. The 0.1 $\mu{s}$ value coincides with the time step used in the FOM calculation. 

Figure~\ref{rom_err_vs_dt} presents the reconstruction error for all three ROM methods with results plotted for mode number increments of 20 POD modes. Using more than 120 modes consistently results in stable ROMs for all three methods whereas less than 100 modes lead to unstable ROMs. Using the reconstruction error as the sole basis for evaluation, there appears to be little reason for preferring one method over the other. In particular, the expected improvement with the LSPG method is not realized in the present reacting flow calculations. As indicated later, this is probably because the major challenges in reacting flow ROMs arise from under-resolved local small-scale dynamics corresponding to flame dispersion and high temperature/species gradients and these completely overwhelm the global, system-level stabilization provided by the LSPG method~\cite{Huang2018July}. Later calculations in which the local errors are controlled show the LSPG method produces a slight improvement over the Galerkin method.

The results in Fig.~\ref{rom_err_vs_dt} also indicate that all three methods are quite insensitive to the time step. Changing the time-step has a very minor impact on the LSPG solutions and an even smaller effect on the Galerkin predictions. Carlberg et al.~\cite{Carlberg2017} have identified a potential sensitivity to time step in the LSPG method, but it does not appear to be a significant issue in the present work. Setting the ROM time step equal to the FOM time step gives essentially identical reconstruction errors for all three methods. The minor differences in the LSPG method show that the ROM solutions with the larger time step (0.2 $\mu{s}$) are marginally less accurate, but that the (0.05 $\mu{s}$) and (0.1 $\mu{s}$) solutions are essentially identical. An issue that may be a factor in the present results is that implicit time stepping (for either the Galerkin or LSPG methods) requires the calculation of Jacobians from the ROM-reconstructed solutions. It is possible that accuracy may be impacted not only by the truncation of POD modes (i.e., small-scale dynamics) but also by errors in the computed Jacobians. In general, Jacobian calculations are more challenging for reacting flow simulations with stiff kinetics, and these errors may escalate in regions where the solution strays from the original calculation. This issue is highlighted later as a potential issue in the implicit Galerkin results. The overall insensitivity to time step indicates the capability of ROMs in providing consistent modeling fidelity, even when larger time steps are used, thereby resulting in improved computational efficiency.

\begin{figure}
	\centering
	\includegraphics[width=1.0\textwidth]{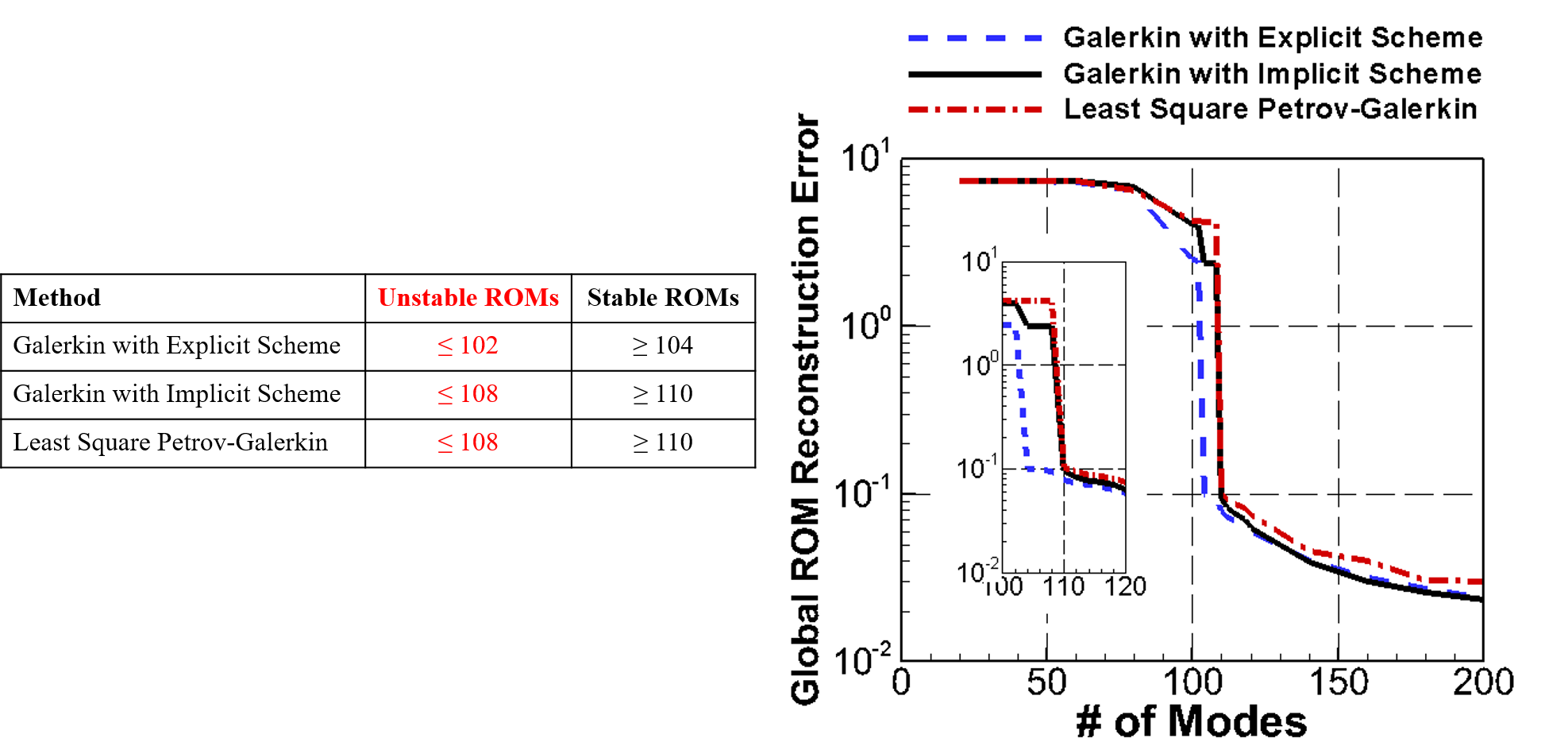}
	\caption{Global ROM reconstruction error comparisons between different projection-based ROM methods with computing time step, $\Delta{t}$ = 0.1 $\mu{s}$, for the 2D benchmark reacting flow simulation.}\label{rom_err_vs_scheme} 
\end{figure}

To investigate the effect of the number of modes on stable ROMs in more detail, Fig.~\ref{rom_err_vs_scheme} shows reconstruction results for mode numbers between 100 and 120 POD modes based upon a mode number increment of two. For consistency with the FOM, the time step in all ROM calculations has been selected as $\Delta{t}$ = 0.1 $\mu{s}$. As highlighted in the zoomed-in view in Fig.~\ref{rom_err_vs_scheme}, at least 104 POD modes are required to generate a stable ROM for the Galerkin method with explicit time marching and at least 110 modes for both the LSPG and Galerkin methods with implicit time marching. More importantly, using two fewer POD modes for any of the three methods leads to unstable ROMs indicating a strong sensitivity to the number of POD modes.

These observations naturally lead to the following questions: 
\begin{enumerate}
    \item Why does a difference of two POD modes have a significant impact on ROM stability?
    \item What are the underlying factors that determine ROM stabilization in reacting flow simulations?
\end{enumerate}
To address these issues, more detailed investigations are presented in the following sections. In particular they focus on three potential factors that may impact numerical stability of the ROMs: 1) conservation law preservation; 2) loss of dissipation, and 3) the existence of unphysical local phenomena.

\subsection{Conservation Law Preservation}
\label{results-ConsvLaw}

\begin{figure}
	\centering
	\includegraphics[width=1.0\textwidth]{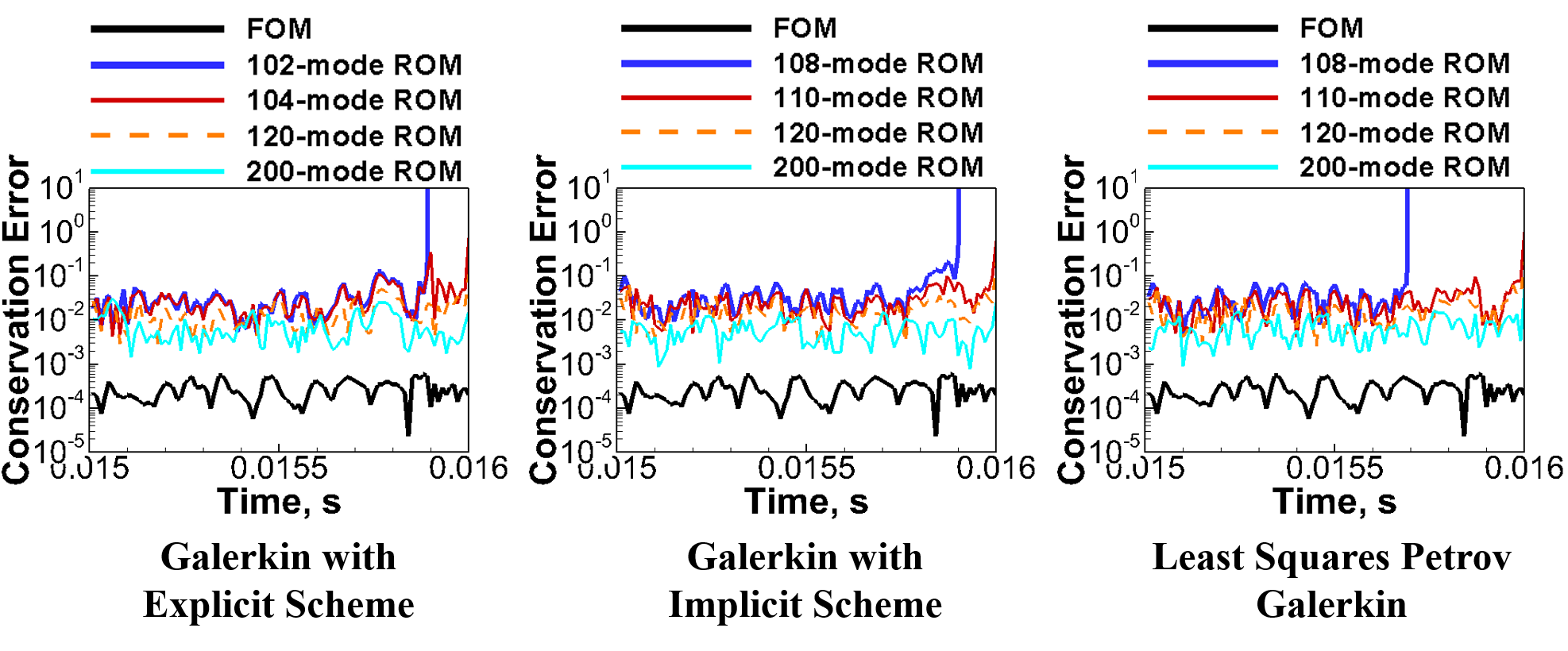}
	\caption{Global primary conservation error comparisons between different projection-based ROM methods with varied computing time steps for the reacting flow simulations.}\label{rom_prim_consv_err} 
\end{figure}

Carlberg et al.~\cite{Carlberg2018Consv} have pointed out that neither Galerkin nor LSPG projection methods implicitly preserve fluid conservation laws for finite-volume models and have demonstrated that ROM stability can be improved by explicitly enforcing conservation in the computed ROMs. Therefore, as a first evaluation of potential contributing factors to ROM instabilities, the degree to which primary and secondary conservation laws are violated in the ROM solutions is assessed. As an indicator of primary conservation law preservation, we define a \emph{global conservation error} as:
\begin{equation}
    \scalebox{0.98}{$\text{Global Conservation Error}(n_p)=\frac{1}{n_{eq}}\sum^{n_{eq}}_{j=1}\left(\frac{\int_V\frac{\partial{Q_j}}{\partial{t}}dV+\sum^{N_{BC}}_{k=1}\int_{\partial\Omega_k}(F-F_v)_j\cdot{dS}-\int_V{H_j}dV}{\sum^{N_{Inflow BC}}_{k=1}\int_{\partial\Omega_k}(F-F_v)_j\cdot{dS}}\right)$}
    \label{rom:prim_consv_err}
\end{equation}
where $j$ represents the particular equation. In determining the conservation error, the temporal storage, flux and source terms for each governing equation (continuity, momentum, energy and species) are computed for each cell at each time step and globally integrated to document the conservation errors. The errors are then normalized by the fluxes at the inflow boundaries and summed over all governing equations. The temporal evolution of the conservation errors in the LSPG and the explicit and implicit Galerkin ROM solutions are then compared with each other and the errors in the FOM. The results are summarized in Fig.~\ref{rom_prim_consv_err}. 

Conservation errors for each ROM method are presented for four different numbers of modes: the highest unstable and lowest stable number of modes for each method, plus 120 and 200 modes for all three methods. As noted in Fig.~\ref{rom_err_vs_scheme}, the lowest-unstable/highest-stable number of modes for the Galerkin explicit method is 108 and 110, while the corresponding pair for the Galerkin implicit and LSPG implicit methods is 102 and 104. The 120-mode case provides a common assessment of all three methods at a number of modes where all methods perform well. The 200-mode case corresponds to the inclusion of the entire POD database in the ROM. 

The FOM solutions exhibit conservation errors on the order of $10^{-4}$ (Fig.~\ref{rom_prim_consv_err}), a level \st{considered} that is typically considered sufficient for reacting flow simulations. The conservation errors in the ROM solutions are more than an order of magnitude larger than the FOM errors, but in all cases, the error is decreased as additional POD modes are included. Even though the primary conservation errors in the ROMs are larger than those in the FOM, the stability of the ROMs does not appear to be related to an increase in conservation error, especially comparing the mode numbers straddling the unstable/stable boundary. For example, the 102-mode ROM from the explicit time marching Galerkin solution exhibits a very similar error evolution as the 104-mode ROM up until the solution blows up ($t \sim$ 0.0159$s$). Similar trends are observed for the Galerkin implicit and LSPG methods although the solution blows up at somewhat different times for each method. It is of particular note that the errors remain nominally constant until blow-up occurs and then exhibit large increases.

\begin{figure}
	\centering
	\includegraphics[width=1.0\textwidth]{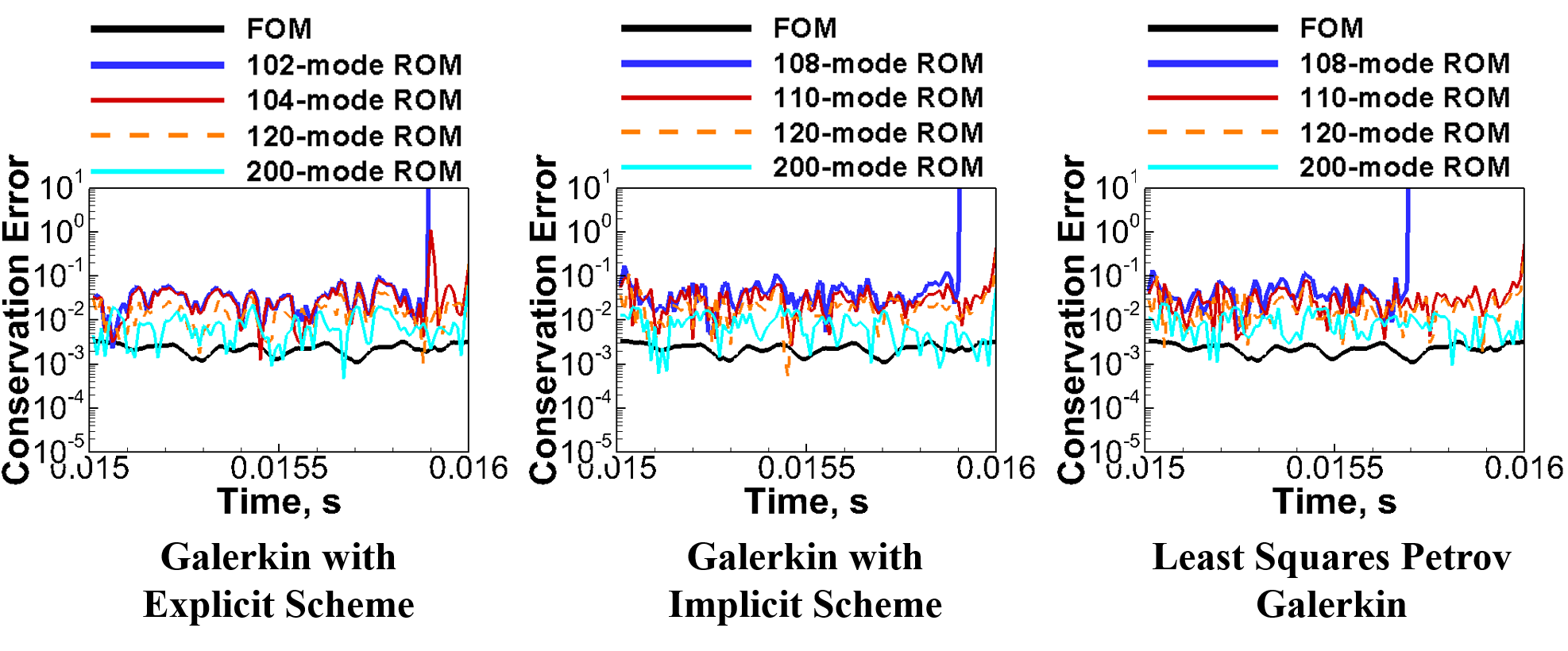}
	\caption{Global secondary conservation error comparisons between different projection-based ROM methods for the reacting flow simulations.}\label{rom_2nd_consv_err} 
\end{figure}

The preservation of secondary conservation is also of importance in numerical schemes~\cite{Tadmor2003} and its importance in ROM stability has been emphasized by Afkham et al.~\cite{Afkham2018}. Both the FOM and ROM solutions are based on the primary conservation laws, but secondary conservation (kinetic energy, $k$, and entropy, $s$) is evaluated by means of the following,
\begin{equation}
    \textbf{Kinetic Energy: } \frac{D\rho{k}}{Dt}=u_j\frac{\rho{u_j}}{Dt}-k\frac{D\rho}{Dt}
    \label{rom:2nd_consv_err-k}
\end{equation}
\begin{equation}
    \textbf{Entropy: } \frac{D\rho{s}}{Dt}=\frac{1}{T}\left(\frac{D(\rho{h^0}-p)}{Dt}-\frac{D\rho{k}}{Dt}-\sum_k(h_k-T{s}_k)\frac{D\rho{Y}_k}{Dt}\right)
    \label{rom:2nd_consv_err-s}
\end{equation}
where $k=\frac{1}{2}u^2_i$. Specifically, kinetic energy ($k$) conservation is evaluated using mass and momentum conservation while entropy ($s$) conservation is obtained from energy, kinetic energy and species conservation. Similar to primary conservation errors, secondary conservation errors are averaged for kinetic energy and entropy for both FOM and ROMs and compared in Fig.~\ref{rom_2nd_consv_err}. The FOM shows a one-order of magnitude larger error in secondary conservation than primary while the ROMs exhibit similar errors in both but still somewhat larger errors than in the FOM. Again, there appears to be no direct connection between secondary conservation errors and ROM stability, suggesting that conservation errors are not a dominant contributor to the stability issues observed in Figs.~\ref{rom_err_vs_dt} and~\ref{rom_err_vs_scheme}.

\subsection{Loss of Dissipation}
\label{results-LossOfDiss}

\begin{figure}
	\centering
	\includegraphics[width=1.0\textwidth]{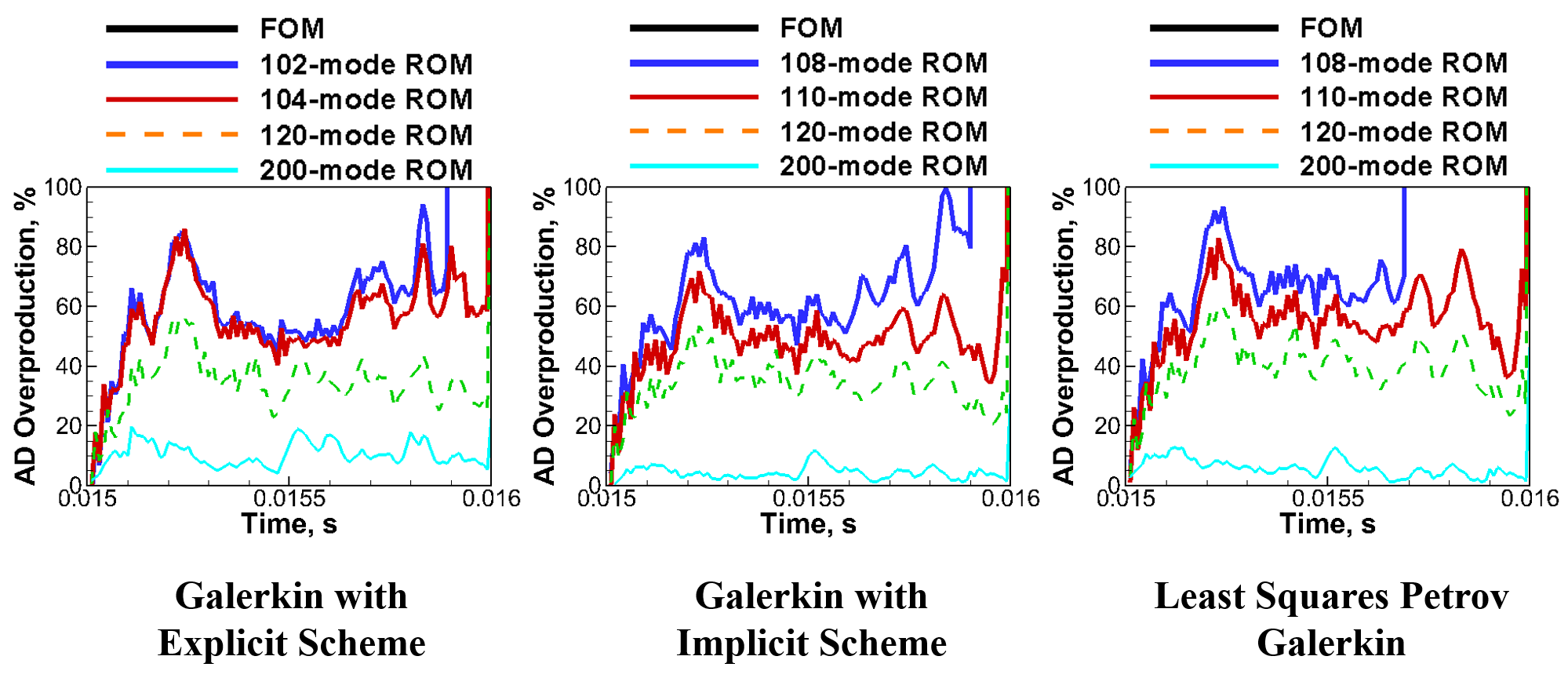}
	\caption{Evaluation of artificial dissipation (AD) overproduction between different projection-based ROM methods for the 2D benchmark reacting flow simulation.}\label{rom_dissipation} 
\end{figure}

A second potential contributor to ROM instability is that the amount of dissipation in the ROM ODE system may be less than that in the FOM PDE because high-energy modes that contribute to dissipation are truncated~\cite{Bergmann2009}. Introducing additional dissipation via a closure model~\cite{Bergmann2009} or numerical dissipation~\cite{Lucia2003} has been demonstrated to improve ROM stability in non-reacting flow problems. Accordingly, we compare the level of artificial dissipation (AD) in the ROM solutions with that in the FOM.
\begin{equation}
    \text{Fractional change of AD}=\frac{\int_V{|\nabla\cdot{\vec{F}_d}|_{ROM}-|\nabla\cdot{\vec{F}_d}|_{FOM}}dV}{\int_V{|\nabla\cdot{\vec{F}_d}|_{FOM}}dV}\times{100\%}
    \label{rom:loss_of_diss}
\end{equation}
where $\nabla\cdot{\vec{F}_d}=\frac{1}{\Delta{V}}\sum^{N_{face}}_{n=1}\left[\frac{1}{2}\Gamma_p|\Gamma^{-1}_p{A_p}|(\hat{Q}_{p,R}-\hat{Q}_{p,L})\right]\cdot\vec{A}_n$ based upon a Roe-finite-volume, spatial discretization. A positive fractional change implies the AD in the ROM is larger than that in the FOM, while a negative value implies the converse. As with the conservation-error comparisons, the level of artificial dissipation in the ROMs is computed for different numbers of modes for the three methods. The results are shown in Fig.~\ref{rom_dissipation}.

The first observation from Fig.~\ref{rom_dissipation} is that the relative change in AD is positive for all ROMs. For the present calculations, all ROMs contain more artificial dissipation than the FOM (As noted later, the increased AD in the ROM solutions may be a result of the steeper gradients in the ROM solutions compared to  those in the FOM solutions). As additional POD modes are included, the overproduction of AD in the ROMs gets smaller, but even the 200-mode ROM contains some 10 - 15\% more AD than the FOM, with the Galerkin explicit being slightly higher than the LSPG and Galerkin implicit methods. Second, for all three methods, stable ROMs are obtained even for AD overproduction levels up to 80\%. More importantly, the level of AD in the unstable ROMs (the blue line in each plot) does not behave significantly different from that of the stable ROMs in which only two more POD modes are included. Though an excess accumulation of AD is present in the Galerkin implicit scheme before it blows up, the ROMs of the other two methods blow up at a time where the AD is near a local minimum. The same assessment has been performed for viscous dissipation (not shown) and leads to similar conclusions. 

In summary, there appear to be no distinguishable connection between ROM blow-up and the level of AD or the level of primary or secondary conservation. A remaining possibility is that instabilities in the ROM solutions arise because the ROM solutions generate unphysical local phenomena. This issue is discussed in the following section.

\subsection{Unphysical Local Phenomena}
\label{results-local}

\begin{figure}
	\centering
	\includegraphics[width=1.0\textwidth]{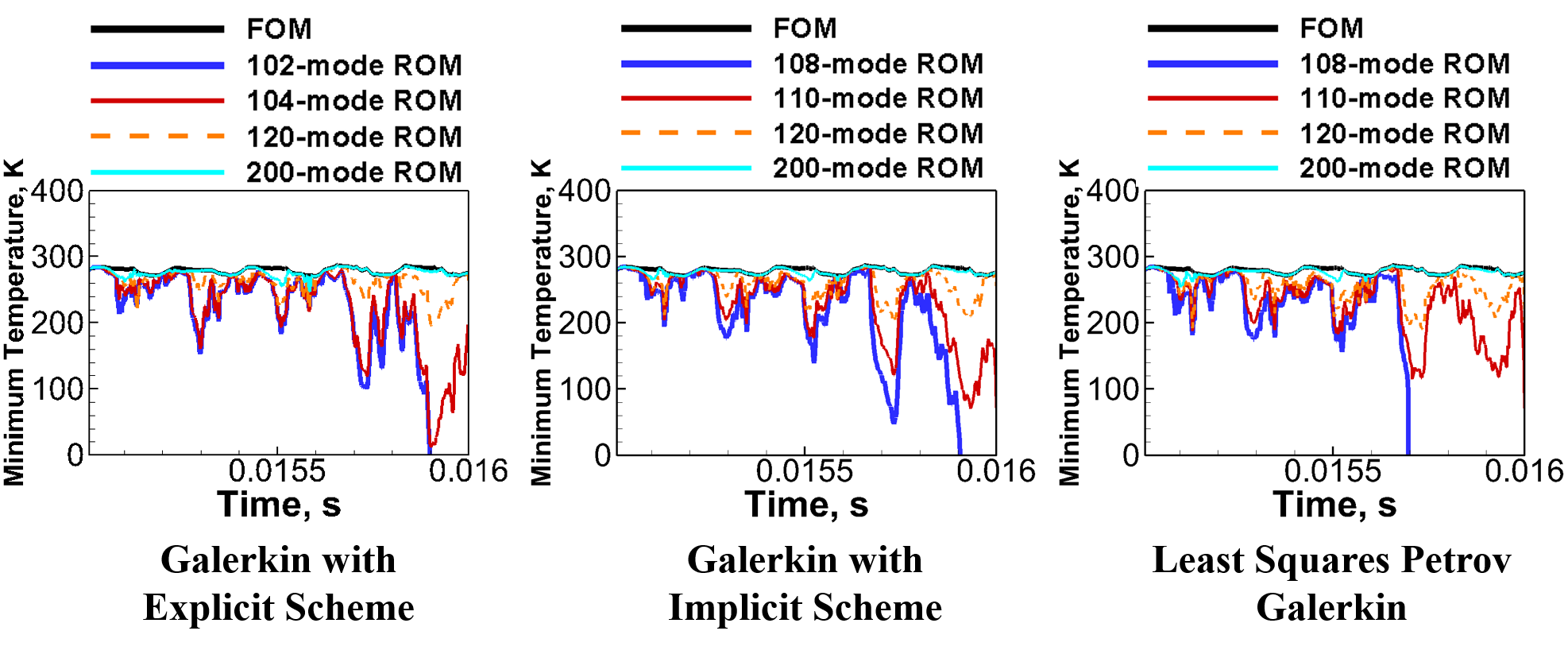}
	\caption{Global minimum temperature values for projection-based ROM methods for the reacting flow simulation.}\label{rom_LocalMinTemp} 
\end{figure}

Thus far, the ROM stability contributor assessment is based on evaluations of global properties, similar to most ROM stabilization methods in the literature including LSPG. In this section, the focus of the assessment is on local phenomena. Due to the large number of degrees of freedom in the FOM and ROM solutions (10,000 snapshots $\times$ 40k cells), the minimum and maximum values of the primitive variables (pressure, velocities, temperature and species) are first investigated for each snapshot from the ROMs to help identify local unphysical phenomena. These investigations draw our attention to the temporal evolution of minimum temperature in the domain at each time step as presented in Fig.~\ref{rom_LocalMinTemp} for the three different ROM methods. The minimum temperatures in the FOM solution are also included for reference. 

An immediate observation from Fig.~\ref{rom_LocalMinTemp}, is that all three ROM methods contain temperatures that are far below the temperature of the incoming fluids whereas the minima in the FOM remains at essentially the incoming temperature. The full 200-mode ROMs show reasonable temperature levels that correlate well with the FOM solution, but as the number of modes in any of the three ROM methods is decreased, the minimum temperature decreases to unphysical values that are substantially below the incoming fluid temperature. The minimum temperature fluctuates in time in response to the imposed periodic forcing at the downstream end, but decreases consistently as time goes on. Most significantly, the unstable ROM for each of the three methods (102 modes for Galerkin with explicit time marching and 108 modes for the LSPG and Galerkin implicit methods) decrease to absolute zero (or below). This ultimately unphysical temperature is clearly the cause for the abrupt solution departure. Unphysical temperatures prove to be the major reason unstable ROMs are generated in this reacting flow problem. 

Even before negative temperatures are reached, the presence of unphysically low temperatures can cause other problems in the ROM calculations. Unphysical temperatures lead to errors in density, sound speed and thermal properties, which can give rise to ill-conditioned Jacobian matrices in the ROM calculations causing further errors. Of particular importance in reacting flow problems is the strong interaction between the local temperature and the highly nonlinear Arrhenius terms that appear in the source terms in the species equations.  A small error in temperature generates a much larger error in the local reaction rate, thus exacerbating the error.  The presence of unphysical local temperatures addresses both of the questions proposed at the end of Section~\ref{results-RomReconErr} and explains why LSPG does not provide significant stability improvements in the resulting ROMs. The unphysical temperatures appear in both Galerkin and LSPG solutions so the global stability improvements of the LSPG method are swamped by these local errors.

\begin{figure}
	\centering
	\includegraphics[width=0.7\textwidth]{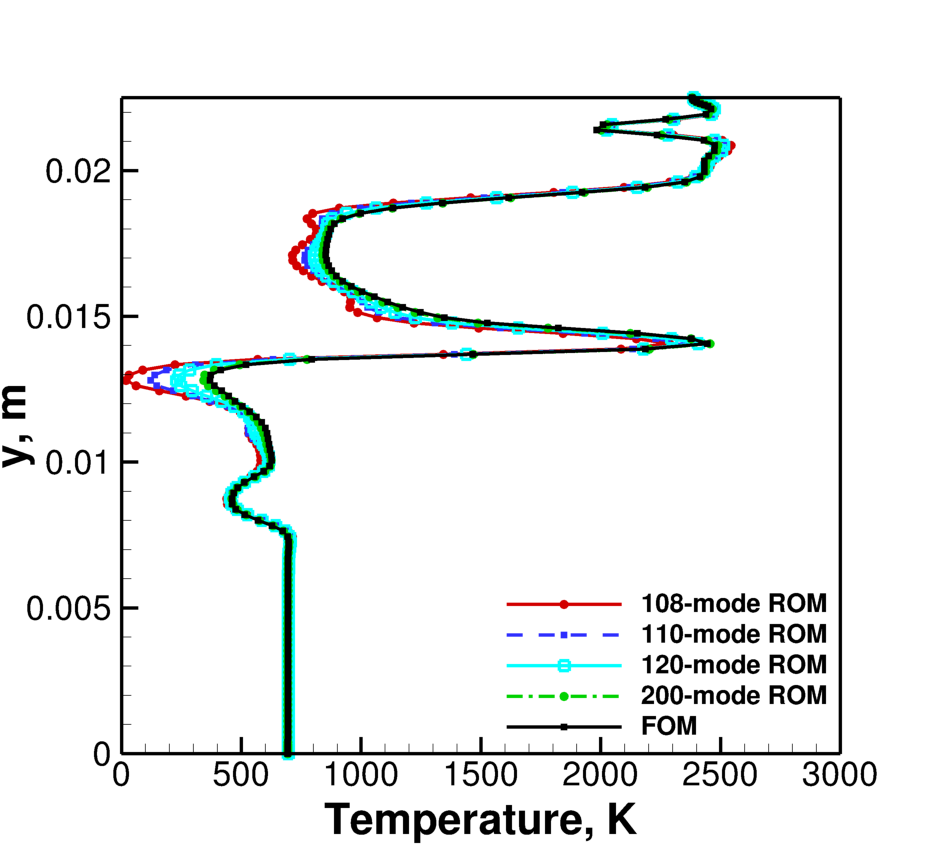}
	\caption{Spanwise temperature profile comparisons between FOM and ROMs at representative streamwise location (x = 0.009m in Fig.~\ref{geometry}) and blow-up time instance (identified from Fig.~\ref{rom_LocalMinTemp}) for Galerkin ROM with implicit scheme.}\label{rom_TempProfile} 
\end{figure}

To provide more insight into the source of these unphysical temperatures, instantaneous spanwise temperature profiles for the FOM and ROM solutions are compared at the approximate streamwise location where the minimum temperature is observed. The temperature excursions for all three ROM methods are similar but for brevity, only results from Galerkin ROMs with implicit time stepping are shown in Fig.~\ref{rom_TempProfile}. The axial location shown is x = 0.009m (see Fig.~\ref{geometry}) and the time is chosen to coincide with the time step at which the 108-mode ROM solution in Fig.~\ref{rom_LocalMinTemp} ($\sim$ 0.0159s) diverges. As shown in Fig.~\ref{rom_TempProfile}, it can be readily seen that the triggering of such unphysical temperature values is the result of a local undershoot in the region adjacent to the sharp temperature gradient from the combustion. In this region, the temperature rises abruptly from 300 to 2500K within 1mm. The over/under-shoots in the ROM solution are a consequence of Gibbs phenomena arising from using a limited number of POD modes for ROM construction. As more POD modes are added, oscillations in the ROM solutions decrease consistently, essentially disappearing when all modes are included. It is these drastic fluctuations in temperature that cause the dissipation in Fig.~\ref{rom_dissipation} to be overproduced in the ROMs. Gibbs phenomena trigger both local over- and under-shoots, increase spatial gradients in all variables, and produce increased dissipation according to Eq.~\ref{rom:loss_of_diss}. 

To summarize, the stability issues observed in ROM of the present reacting flow problem can be primarily attributed to the triggering of unphysical local phenomena in which temperatures drop to extremely low values as a consequence of Gibbs phenomena near sharp temperature gradients. While undershoots that reach absolute zero are absolutely detrimental, overshoots in reacting flow regions can lead to dramatic increases in reaction rates that also lead to serious accuracy problems. POD-based ROM solutions are analogous to standard spectral methods, which are well recognized as being susceptible to Gibbs phenomena in the presence sharp gradients. These temperature issues are further addressed in the following section in which ROM predictions of non-reacting flows with and without steep temperature gradients are considered.

\subsection{Temperature Fluctuations in Non-reacting Flows}
\label{results-non-reacting}

As a means of distinguishing the effects of steep temperature gradients from the effects of chemical reactions, we briefly consider a set of non-reacting flows. For consistency with the reacting flow problem, three sets of FOM simulations of non-reacting flow are performed using the configuration in Fig.~\ref{geometry}. The incoming fluids contain the same species concentrations in the $T_1$ and $T_2$ streams as the reacting flow case but three different temperature ratios are considered in the two streams. In all cases, the fluid in stream $T_1$ is fixed at 300K but the temperature ratios are chosen as $T_2$ / $T_1$ = 1.0, 3.0 and 4.0) to investigate the effects of temperature variations. Combustion is omitted by setting the pre-exponential factor to zero.

\begin{figure}
	\centering
	\includegraphics[width=1.0\textwidth]{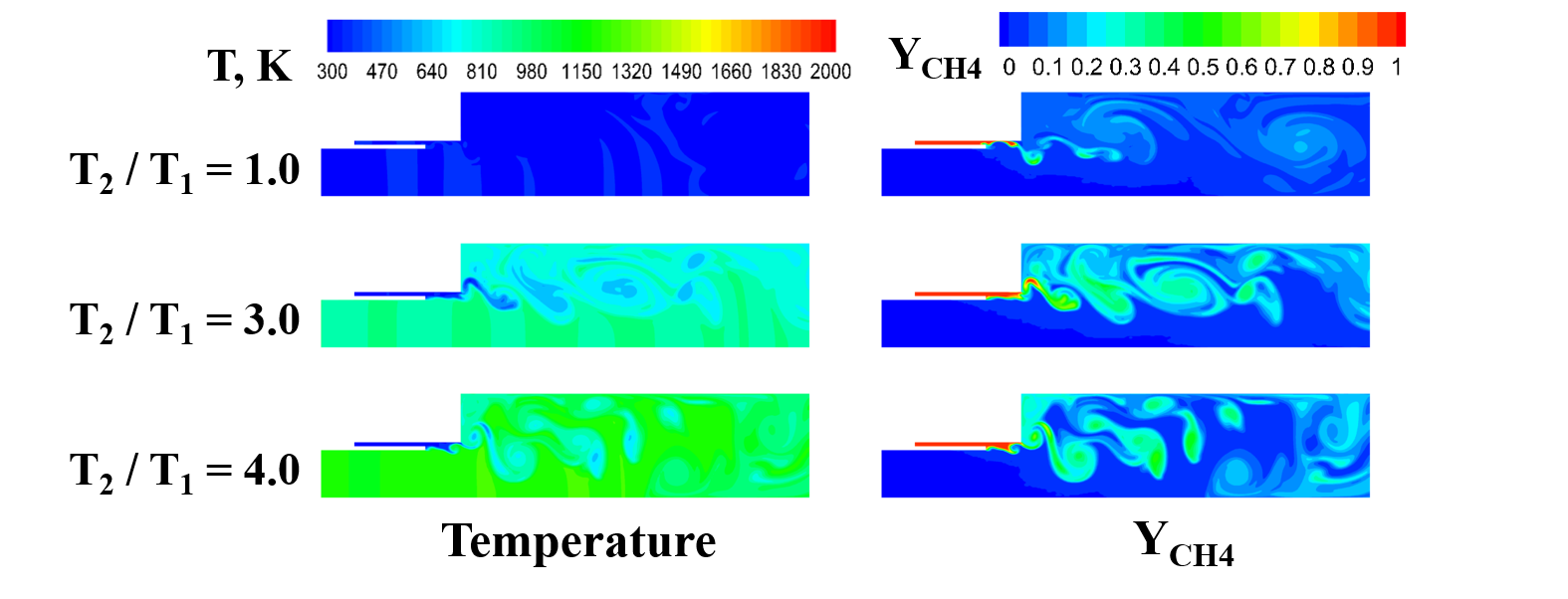}
	\caption{Representative instantaneous snapshots comparisons for 2D benchmark non-reacting flow simulations with varied temperature ratios.}\label{non_reacting_snapshots} 
\end{figure}

Instantaneous snapshots from the three FOM solutions are compared in Fig.~\ref{non_reacting_snapshots}. Overall, the representative dynamics of all cases are qualitatively similar to the reacting flow case and show many of the characteristics expected in reacting flow. Each flowfield exhibits small-scale eddies originating within the recess between the exit of the T1 stream and the dump plane. These small eddies are then shed into the combustor and follow the large-scale recirculation zone created by the dump plane, eventually being absorbed into the larger scales.

\begin{figure}
	\centering
	\includegraphics[width=0.6\textwidth]{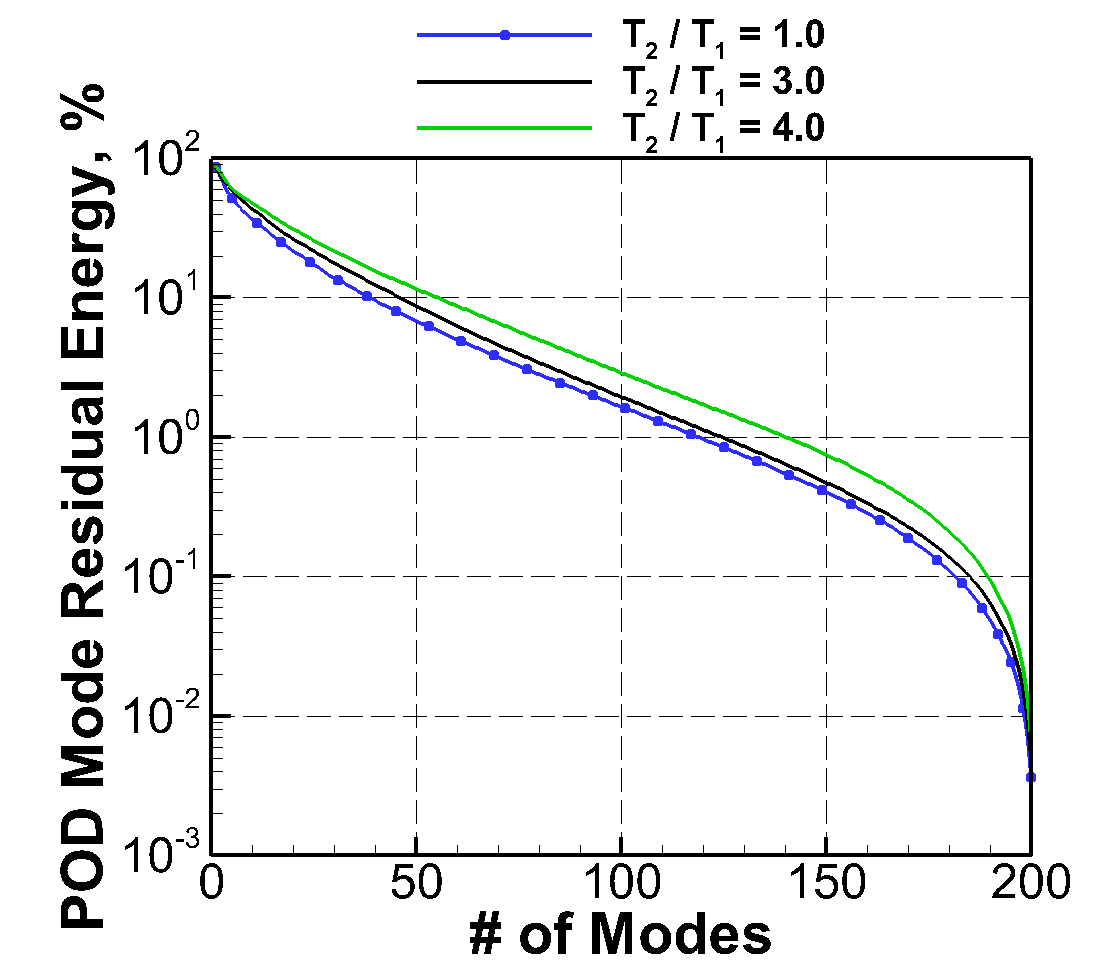}
	\caption{POD residual energy distribution comparisons for 2D benchmark non-reacting flow simulations with varied temperature ratios.}\label{nr_pod_res_energy} 
\end{figure}

The residual energy in the three POD datasets is shown as a function of the number of modes in Fig.~\ref{nr_pod_res_energy}. Comparison with Fig.~\ref{pod_res_energy} reveals that the amount of information omitted by the POD representation for any particular number of modes is essentially the same as in the reacting flow case, independent of the temperature ratio in the non-reacting flow solution with only modest differences between the four datasets.

\begin{figure}
	\centering
	\includegraphics[width=0.6\textwidth]{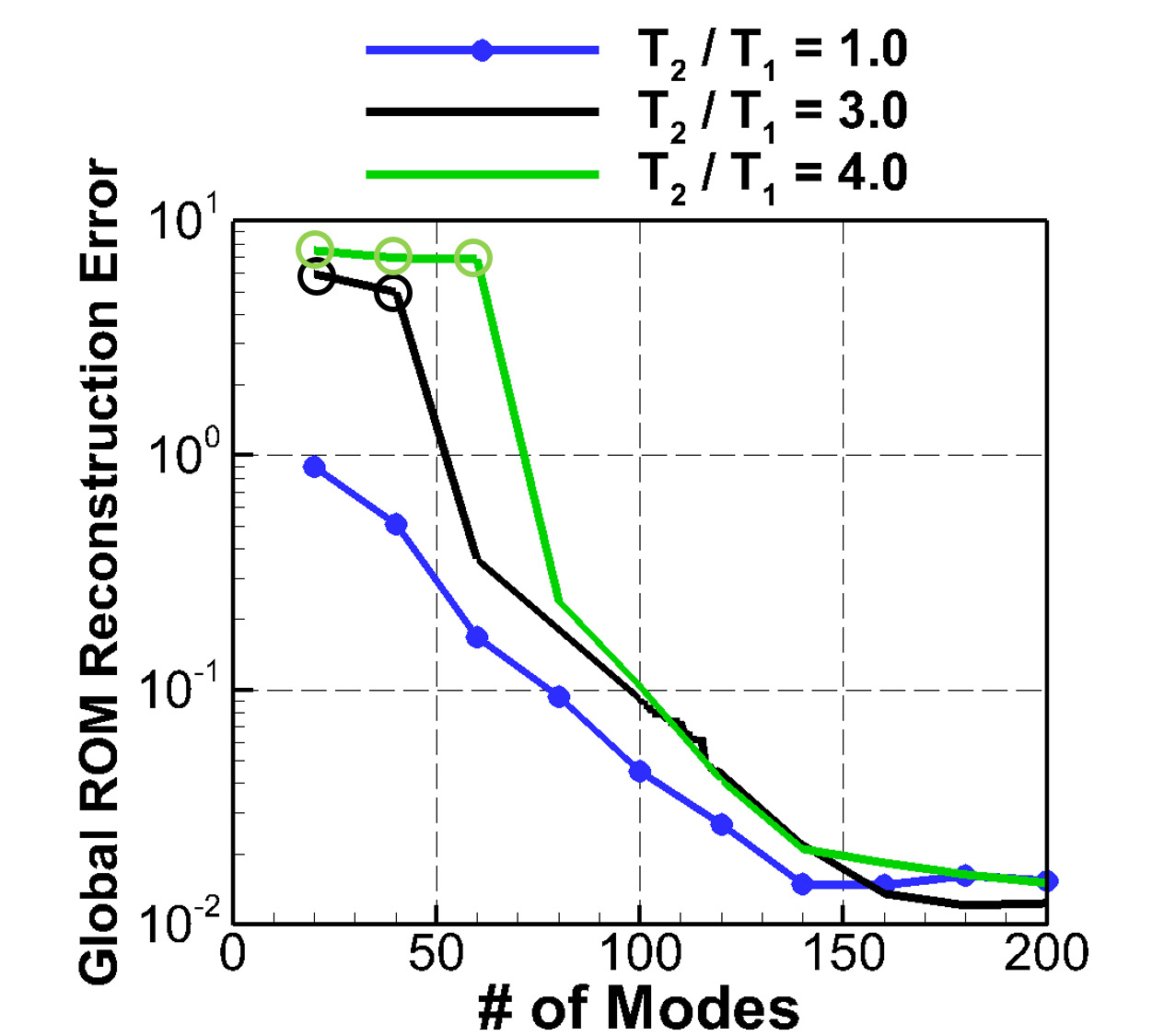}
	\caption{Global ROM reconstruction error comparisons for the 2D benchmark non-reacting flow simulation with varied temperature ratios (unstable ROMs marked in hollow symbols).}\label{nr_recon_err_vs_Tratio} 
\end{figure}

ROM reconstruction errors are plotted as a function of the number of POD modes in Fig.~\ref{nr_recon_err_vs_Tratio}  for all three temperature ratios. For the $T_2$ / $T_1$ =1.0 case, continuous monotone convergence from 20 to 200 modes is observed and all generated ROMs are stable. The errors in the two non-unity temperature ratio cases, however, are very large when a small number of modes is included but drop abruptly after 60 modes in the $T_2$ / $T_1$ = 3.0 case and after 80 modes in the $T_2$ / $T_1$ = 4.0 case. ROMs constructed with less than these numbers of modes are unstable for both cases. Increasing the temperature ratio requires more POD modes to reach a stable ROM and makes it more challenging to generate robust ROMs. Comparison with Fig.~\ref{rom_err_vs_dt} shows that the $T_2$ / $T_1$ = 3.0 and 4.0 cases are highly analogous to the reacting flow cases. Inspection of these non-reacting flows again shows strong temperature over-and under-shoots near steep temperature gradients. The complexities generated by steep temperature gradients is one of main challenges in ROM development for reacting flow simulations.

\subsection{Implementation of Temperature Constraints}
\label{results-temp-constraints}

Both the reacting and non-reacting solutions suggest that the major cause of ROM stability arises because of Gibbs phenomena near steep temperature gradients leading to unphysical temperatures that are not present in the CFD database. As a verification that the temperature excursions are the source of instability, simple temperature constraints are imposed on the ROM calculations. After each ROM time iteration, the temperatures are constrained in between 250 and 2750K. The lower bound of the constraints, 250K, is determined by the cold reactant temperature, 50K below the incoming temperature of 300K while the higher bound, 2750K, is 50K above the adiabatic temperature of the flame of 2700K. Temperature constraints are also imposed for ROMs of the two high temperature ratio non-reacting flow cases in Fig.~\ref{nr_recon_err_vs_Tratio} with the same lower bound but 950K and 1250K for the higher bounds for $T_2$ / $T_1$ = 3.0 and $T_2$ / $T_1$ = 4.0 respectively. (Adding constraints on the $T_2$ / $T_1$ = 1.0 case has no impact as the limits are never exceeded.) 

\begin{figure}
   \subfloat[Non-reacting Flow\label{rom_cmp_wthwoTempConstraints:nr}]{\includegraphics[width=0.9\textwidth]{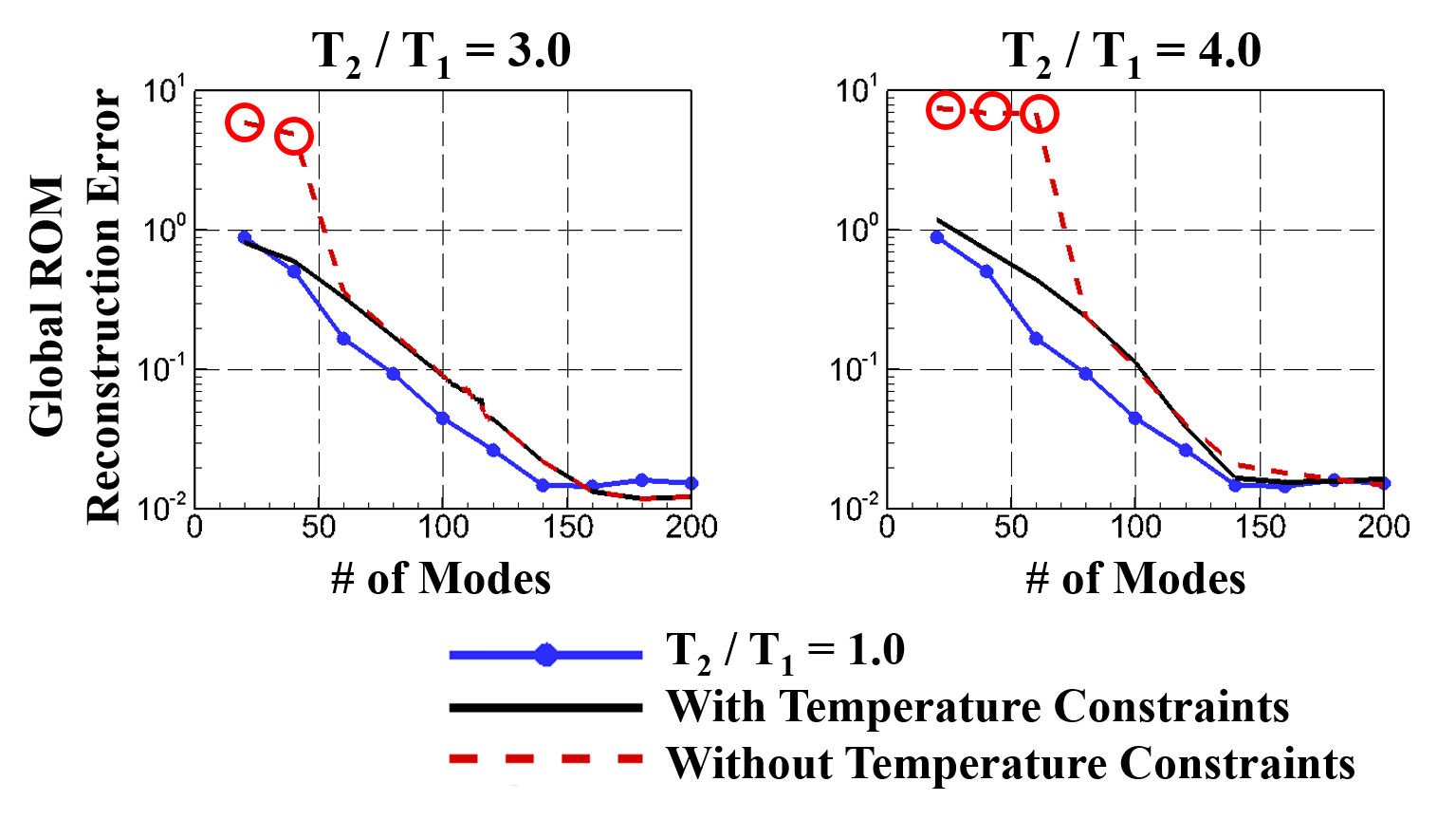}}
   \hfill
   \subfloat[Reacting Flow\label{rom_cmp_wthwoTempConstraints:r}]{\includegraphics[width=\textwidth]{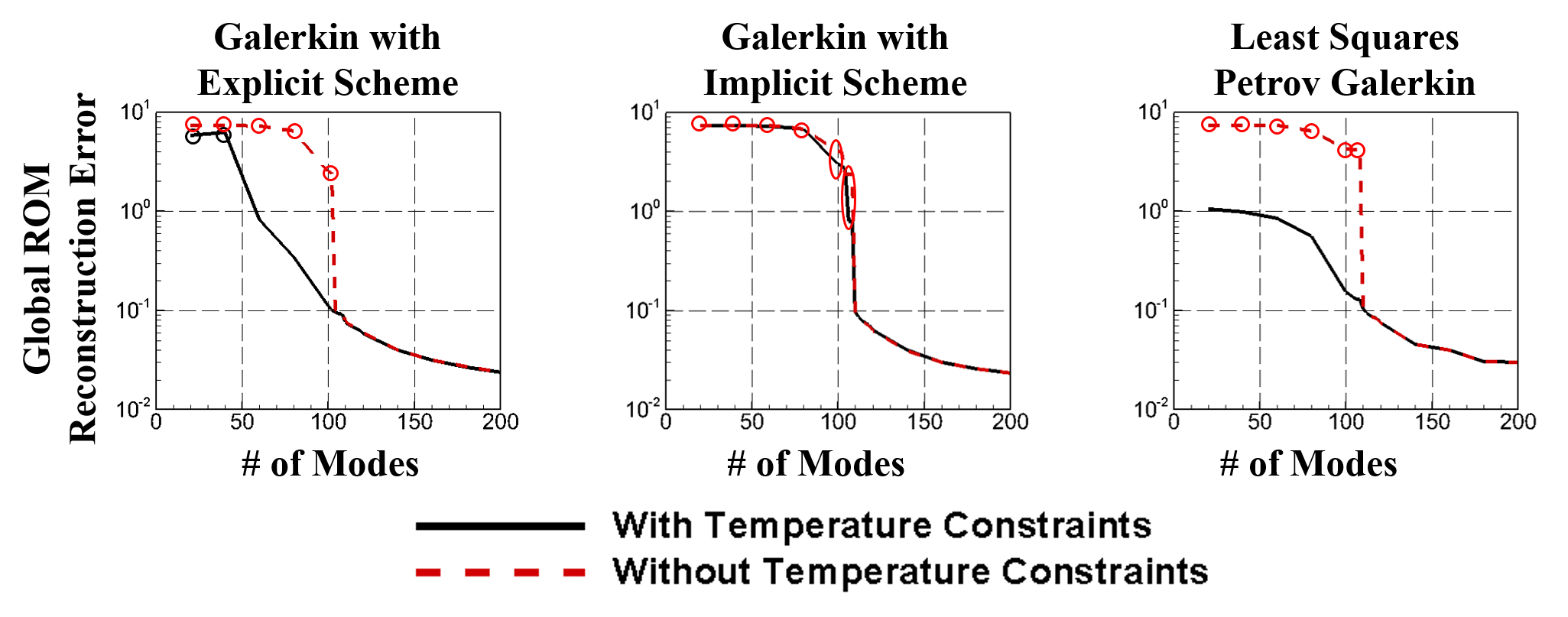}}
   \caption{Global ROM reconstruction error comparisons for the 2D benchmark non-reacting and reacting flow simulations with temperature constraints (unstable ROMs marked in hollow symbols).}\label{rom_cmp_wthwoTempConstraints} 
\end{figure}

The global ROM reconstruction errors with temperature constraints imposed are compared in Fig.~\ref{rom_cmp_wthwoTempConstraints} with solutions without the constraints. The plots at the top compare the non-reacting cases. Results are shown for $T_2$ / $T_1$ = 3.0 and 4.0 with and without constraints imposed, and for $T_2$ / $T_1$ = 1.0 (where constraints have no effect). Imposing constraints brings the convergence error for both $T_2$ / $T_1$ = 3.0 and 4.0 nearly into coincidence with that for the $T_2$ / $T_1$ = 1.0 case. The ROM convergence error starts at a stable level and decreases monotonically as additions modes are added. All ROMs generated from 20 to 200 modes are stable and robust.

The plots at the bottom of Fig.~\ref{rom_cmp_wthwoTempConstraints} present ROM convergence for the reacting flow problem for the Galerkin and LSPG methods with and without temperature constraints. The Galerkin explicit and LSPG methods show substantial improvements when temperature constraints are employed but the Galerkin implicit method shows little difference. The effect of temperature constraints is most beneficial for the LSPG results. With temperature constraints, the LSPG method produces stable ROMs for all POD mode numbers from 20 to 200 with nearly monotonic convergence in ROM errors. Without constraints, 110 modes are required for stability. Both Galerkin solutions retain regions of instability when small numbers of POD modes are used. For the Galerkin solution with explicit time marching, stable ROMs can be obtained by including 60 or more POD modes whereas without temperature constraints a minimum of 104 modes is required. The Galerkin solutions with implicit time marching, however, show essentially no improvement with temperature constraints. Implicit Galerkin solutions still require 110 POD modes to obtain stable ROMs. Further analysis of the Galerkin implicit results indicate that even though the existence of extreme local phenomena has been prevented, the mass matrix of the ROM ODEs (similar to $\hat{M}^n$ in Eq.~\ref{rom:ode-lspg}) still exhibits poor numerical properties that lead to numerical divergence during ROM iterations and eventually terminate the calculations. On the other hand, both Galerkin with explicit scheme and LSPG methods produce symmetric mass matrices (identity matrices for the explicit scheme), which in general have better numerical properties. 

\begin{figure}
	\centering
	\includegraphics[width=0.6\textwidth]{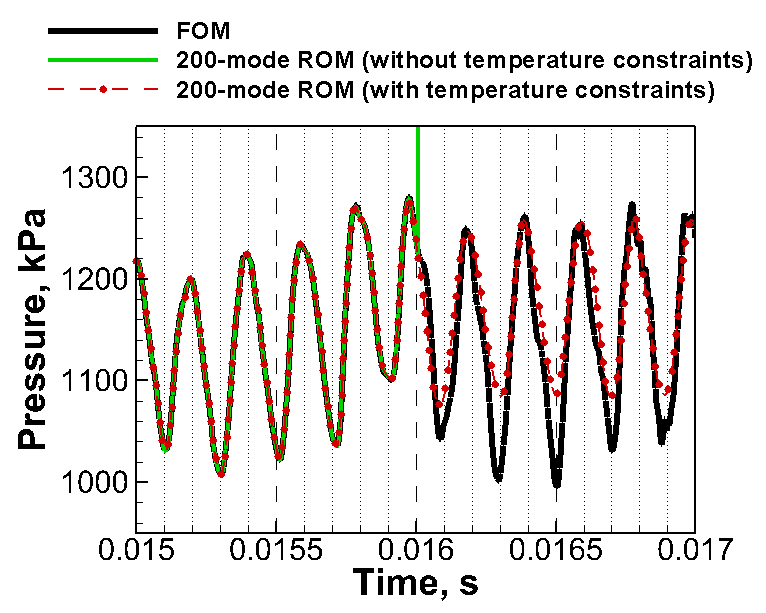}
	\caption{Evaluations of future-state predictive capabilities for LSPG ROM  with and without temperature constraints based on pressure signal taken near the dump plane (x = 0 m in Fig. ~\ref{geometry}).}\label{lspg_future_state} 
\end{figure}

As a final observation, a further benefit of temperature constraints is that they enable the ROMs to be extended to future-state predictions beyond the time interval used for basis construction. Comparisons between the FOM and 200-mode ROMs for reacting flow with and without temperature constraints are shown in Fig.~\ref{lspg_future_state} for the LSPG ROM methods. Future state predictions with the Galerkin methods are similar. Both with and without temperature constraints, the 200 POD modes produce stable and accurate ROMs within the reconstruction interval (0.015 to 0.016s), however, as demonstrated in the pressure signal comparisons in Fig.~\ref{lspg_future_state}, only the ROMs with temperature constraints are able to predict solutions beyond 0.016s. ROMs without constraints blow up shortly after the calculations proceed outside the reconstruction interval. Although the capability of future-state predictions requires further investigations and improvement, as is the case in simpler non-reacting flow problems~\cite{Carlberg2017}. It should also be pointed out that such advancement can also be achieved for Galerkin method as well, both explicit and implicit with the temperature constraints. Similar improvements in future-state predictions were  achieved for the non-reacting flow cases.

\section{Conclusion}

The accuracy, stability and robustness of reduced order models (ROM) for reacting flow applications has been investigated in a detailed manner. The goal is to reveal the prime underlying causes for numerical instability and evaluate corresponding resolutions to improve the robustness of the ROMs. Galerkin and least squares Petrov-Galerkin (LSPG) methods were compared for a representative combustor problem to assess the baseline ROM characteristics in reacting flow simulations. Overall, ROM stability appears to be highly sensitive to the number of POD modes included and the addition of only two modes can switch from unstable to stable ROMs. Attention was focused on investigating three potential contributing factors for ROM stability: 1) conservation law preservation; 2) loss of dissipation; and 3) unphysical local phenomena.

In addressing the first potential reason for ROM sensitivity, preservation of primary and secondary conservation was quantified and investigated for FOM and ROMs. ROMs exhibit conservation errors more than an order of magnitude higher than those in the FOM, but they remain sufficiently small that stable ROMs are still be produced. Further, the level of conservation errors in unstable ROMs is similar to that in stable ones suggesting that the violation of conservation is not a significant contributor to ROM stability. 

Next, the level of artificial dissipation in the ROMs was shown to be consistently larger than in the FOM with increases of 10\% to 80\%. Again, the global dissipation level was observed to be similar in stable and unstable ROMs, and does not appear to be a significant cause of ROM stability. 

Third, attention was directed toward unphysical local phenomena, which strongly suggest that the blow-up of ROMs is driven by the appearance of unphysical temperatures arising from Gibbs phenomena near steep temperature gradients. Similar conclusions were observed for non-reacting flows with steep temperature gradients.

Based on the identification of unphysical temperatures as a major contributor to instability, the ROM temperature fields were clipped to preclude unphysical values. With temperature constraints, both Galerkin and LSPG methods exhibited significant improvements in ROM stability and error convergence. Imposition of temperature constraints was also shown to enable effective future-state prediction. It should be pointed out that the temperature constraints are not adhoc. The lower bound is determined based on the cold reactant temperature while the higher bound is set based on the adiabatic flame temperature. Clearly, more sophisticated limiters  are preferred, but the present results highlight the necessity of controlling sharp gradients.

Although the imposition of temperature constraints yields significant improvement in ROM performance, it does not eliminate the Gibbs phenomena. This is indicative of the deficiency of flux limiters in eliminating Gibbs phenomena in ROMs or the lack of appropriate closure or stabilization to account for the truncated bases. Therefore, more comprehensive improvements can be achieved by developing ROM-specific limiters~\cite{Ray2018}, more systematic closure models ~\cite{Parish2019} and by using  adaptive bases~\cite{peherstorfer2015online,Huang2018Jan} to  eliminate  Gibbs phenomena. Further consideration should be given to the challenges involved in improving the efficiency of the ROM using sparse sampling techniques~\cite{carlberg2013gnat,  Huang2018Jan}. 

\appendix
\section{ Governing Equations for Full Order Model}
\label{appendix:fom_eq}

The full order model computations are carried out with an in-house CFD code, the General Equations and Mesh Solver (GEMS), the capabilities of which has been successfully demonstrated in modeling rocket combustion instabilities~\cite{HarvazinskiPoF}. GEMS solves the conservation equations for mass, momentum, energy and species mass fractions in a coupled fashion,
\begin{equation}
    \frac{\partial{Q}}{\partial{t}}+\nabla \cdot\left(\vec{F}-\vec{F_v}\right)=H,
    \label{fom:governing}
\end{equation}
where $Q$ is the vector of conserved variables defined as, $Q=\left(\begin{array}{cccccc}
        \rho & \rho{u} & \rho{v} & \rho{w} & \rho{h^0-p} & \rho{Y_l}\\
    \end{array}\right)^T$
with $\rho$ representing density, $u$, $v$ and $w$ representing velocity field, $Y_l$ representing the $l^{th}$ species mass fraction and the total enthalpy $h^0$ is defined as, $h^0=h+\frac{1}{2}(u^2_i)=\sum_l{h_l{Y_l}}+\frac{1}{2}(u^2_i)$.

The fluxes have been separated into inviscid, $\vec{F}=F_{i}\vec{i}+F_{j}\vec{j}+F_{k}\vec{k}$ and viscous terms, $\vec{F_v}=F_{v,i}\vec{i}+F_{v,j}\vec{j}+F_{v,k}\vec{k}$. And the three inviscid fluxes are,
\begin{equation}
    F_i = \left(\begin{array}{c}
        \rho{u} \\
        \rho{u^2}+p \\
        \rho{uv} \\
        \rho{uw} \\
        \rho{uh^0} \\
        \rho{uY_l} \\
    \end{array} \right), \; F_j = \left(\begin{array}{c}
        \rho{v} \\
        \rho{uv} \\
        \rho{v^2}+p \\
        \rho{vw} \\
        \rho{vh^0} \\
        \rho{vY_l} \\
    \end{array} \right) \; \text{and} \; F_k = \left(\begin{array}{c}
        \rho{w} \\
        \rho{uw} \\
        \rho{vw}+p \\
        \rho{w^2}+p \\
        \rho{wh^0} \\
        \rho{wY_l} \\
    \end{array} \right)
    \label{fom:inviscid_fluxes}
\end{equation}
The viscous fluxes are,
\begin{equation}
    \scalebox{0.8}{$F_{v,i} = \left(\begin{array}{c}
        0 \\
        \tau_{ii} \\
        \tau_{ji} \\
        \tau_{ki} \\
        u\tau_{ii}+v\tau_{ji}+w\tau_{ki}-q_i \\
        \rho{D_{l}}\frac{\partial{Y_l}}{\partial{x}} \\
    \end{array} \right), \; F_{v,j} = \left(\begin{array}{c}
        0 \\
        \tau_{ij} \\
        \tau_{jj} \\
        \tau_{kj} \\
        u\tau_{ij}+v\tau_{jj}+w\tau_{kj}-q_j \\
        \rho{D_{l}}\frac{\partial{Y_l}}{\partial{y}} \\
    \end{array} \right) \; \text{and} \; F_{v,k} = \left(\begin{array}{c}
        0 \\
        \tau_{ik} \\
        \tau_{jk} \\
        \tau_{kk} \\
        u\tau_{ik}+v\tau_{jk}+w\tau_{kk}-q_k \\
        \rho{D_{l}}\frac{\partial{Y_l}}{\partial{z}} \\
    \end{array} \right)$}
    \label{fom:viscous_fluxes}
\end{equation}
where $D_l$ is defined to be the diffusion of the $l^{th}$  species into the mixture. In practice, this is an approximation used to model the multicomponent diffusion as the binary diffusion of each species into a mixture.

The heat flux in the $i^{th}$ direction, $q_i$, is defined as,
\begin{equation}
    q_i = -K\frac{\partial{T}}{\partial{x_i}}+\rho\sum^N_{l=1}D_l\frac{\partial{Y_l}}{\partial{x_i}}h_l+\mathbf{Q}_{\text{source}}
    \label{fom:heat_flux}
\end{equation}
The three terms in the heat flux represent the heat transfer due to the conduction, species diffusion and heat generation from a volumetric source (e.g. heat radiation or external heat source) respectively.

The shear stress, $\tau$ , is also found in the viscous flux and defined in terms of the molecular viscosity and velocity field,
\begin{equation}
    \tau_{ij} = \mu\left(\frac{\partial{u_i}}{\partial{x_j}}+\frac{\partial{u_j}}{\partial{x_i}}-\frac{2}{3}\frac{\partial{u_m}}{\partial{x_m}}\delta_{ij}\right)
    \label{fom:shear_stress}
\end{equation}

The source term, $H$ includes a single entry for each of the species equations signifying the production or destruction of the $l^{th}$ species, $\dot{\omega}_l$, which is determined by the chemical kinetics~\cite{WestbrookDryer},
\begin{equation}
    H=\left(\begin{array}{cccccc}
        0 & 0 & 0 & 0 & 0 & \dot{\omega}_l\\
    \end{array}\right)^T
    \label{cfd:source_term}
\end{equation}

\section{Investigation on the Effects of Snapshot Selection}
\label{appendix:snapshot_effects}

\begin{figure}
	\centering
	\subfloat[POD residual energy distribution\label{lspg_snap_select:pod}]{\includegraphics[width=0.5\textwidth]{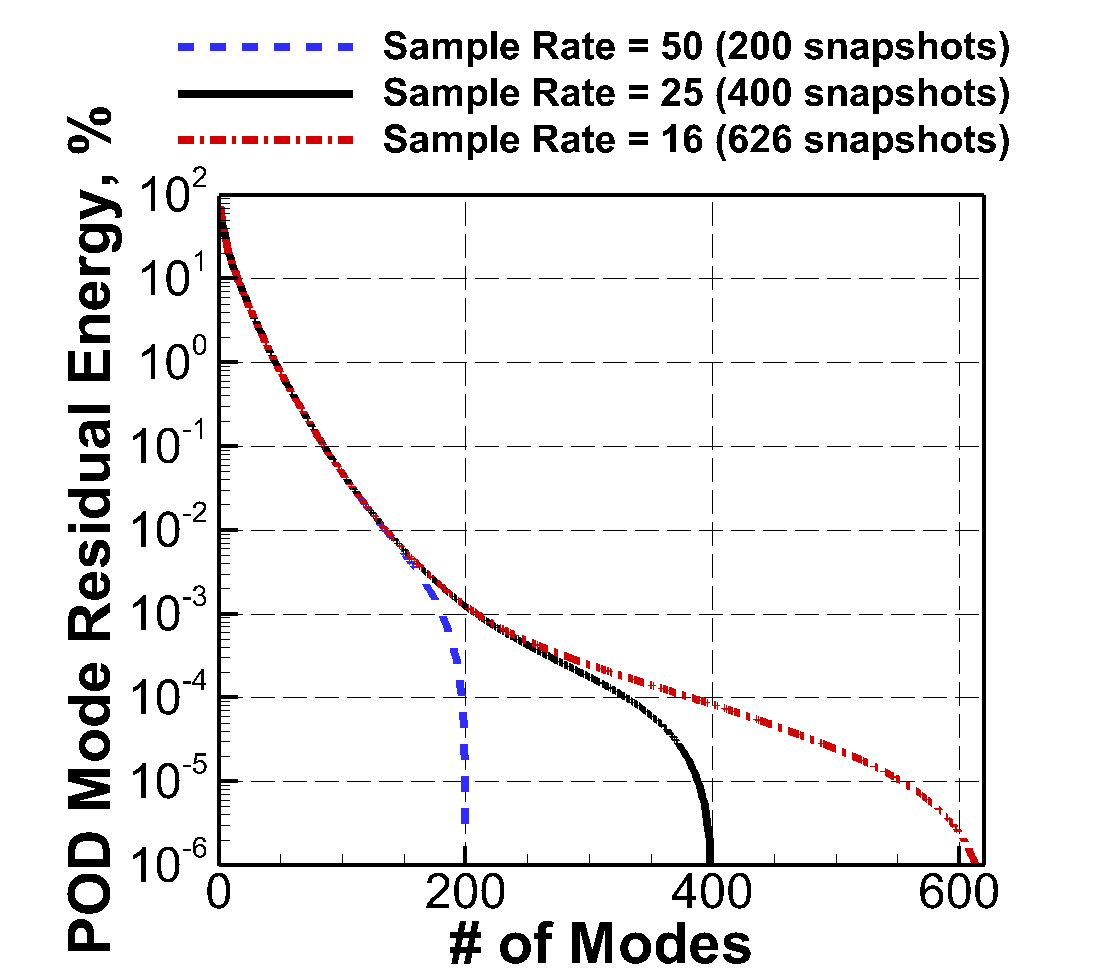}}
	\hfill
	\subfloat[Global ROM reconstruction error (LSPG with temperature constraints)\label{lspg_snap_select:rom}]{\includegraphics[width=0.5\textwidth]{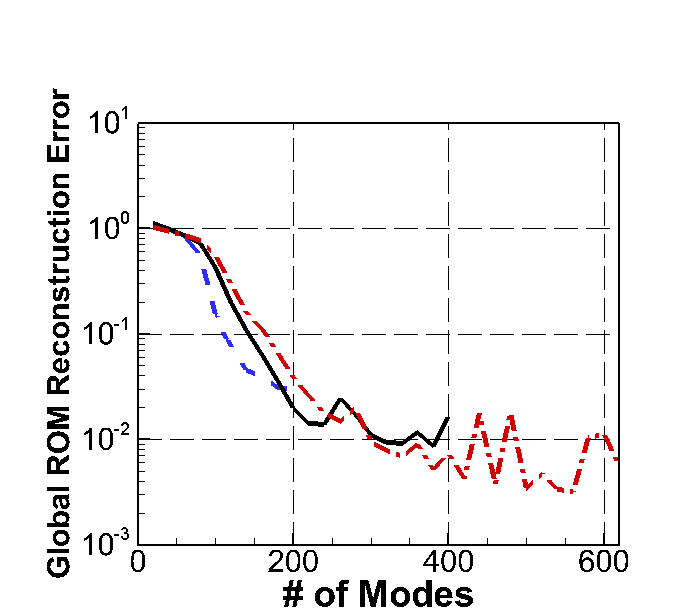}}
	\caption{Parametric investigations on the effects of down-sampling rates.}\label{lspg_snap_select}
\end{figure}

With the improvement on ROM stability via the imposition of temperature constraints as demonstrated in Section~\ref{results-temp-constraints}, further sensitivity studies are carried out to investigate the impact of snapshots selection  on ROM characteristics. Accordingly, the FOM solutions have been down-sampled every 50, 25 and 16 time steps for POD mode generation, which results in 200, 400 and 626 snapshots respectively. It should be noted that all the studies above were carried out using a down-sampling rate of 50 time steps, and by increasing the sampling rates, it is expected that more  high-frequency dynamics will be included in the ROM evolution. The comparisons of the POD residual energy (Eq.~\ref{pod:res_energy}) and the global ROM construction error (Eq.~\ref{rom:reconstr_err}) are shown in Fig.~\ref{lspg_snap_select}. 

 Fig.~\ref{lspg_snap_select:pod} reveals the amount of information omitted by the POD representation for any particular number of modes. It is readily seen that the residual energies for all three sampling rates follow nearly identical decay rates for approximately the first 60 POD modes. Concurrently, it is noted that these first 60 modes contain nearly 90\% of the total energy. The relatively large fraction of modes required to recover 90\% of the energy, in contrast to non-reacting flow problems in the literature, emphasizes  the wide range of important scales in the present problem. When more than 60 modes are included in the POD data set, substantial differences appear between the three sampling rates as more high-frequency dynamics (possibly including numerical noise in FOM solutions) are included with the increase of sampling rates, which usually accumulates in the low-energy POD modes. 

The ROMs are constructed using LSPG with  temperature constraints introduced in Section~\ref{results-temp-constraints}, and the reconstruction errors are compared in Fig.~\ref{lspg_snap_select:rom} for three sampling rates. It can be readily seen that the ROMs from all three sampling rates are stable and exhibit largely monotone error convergence. More importantly, as the sampling rate increases, the accuracy of the ROM reconstruction appears to be consistently improved by at least a factor of 2 to 3, comparing the 200-mode ROMs between sampling rate = 50 and 25 and the 400-mode ROMs between sampling rate = 25 and 16. This is  expected since with a higher sampling rate, there is less information loss, and therefore, the resulting ROMs are anticipated to provide a more accurate representation of the FOM solutions.

\section*{Acknowledgments}
The authors acknowledge  support from the Air Force under the Center of Excellence grant FA9550-17-1-0195, titled Multi-Fidelity Modeling of Rocket Combustor Dynamics. The authors also would like to acknoledge the computing resources provided by the NSF via grant 1531752 MRI: Acquisition of Conflux, A Novel Platform for Data-Driven Computational Physics (Tech. Monitor: Stefan Robila).

\section*{References}
\bibliographystyle{aiaa}
\bibliography{ref.bib}
\end{document}